\documentclass[aps,prb,twocolumn,showpacs]{revtex4}

\usepackage{graphicx}
\usepackage{amsmath}
\usepackage{amssymb}

\begin{document}

\title{Unusual electronic structure in underdoped cuprate superconductors}

\author{Xiang Li$^{1,2}$, Minghuan Zeng$^{3}$, Huaiming Guo$^{4}$, and Shiping
Feng$^{2,1}$}
\thanks{Corresponding author. E-mail: spfeng@bnu.edu.cn}

\affiliation{$^{1}$School of Physics and Astronomy, Beijing Normal University,
Beijing 100875, China}

\affiliation{$^{2}$Department of Physics, Faculty of Arts and Sciences, Beijing
Normal University, Zhuhai 519087, China}

\affiliation{$^{3}$College of Physics, Chongqing University, Chongqing 401331, China}

\affiliation{$^{4}$School of Physics, Beihang University, Beijing 100191, China}

\begin{abstract}
The underdoped cuprate superconductors are characterized by the opening of the
normal-state pseudogap, while such an aspect of the normal-state pseudogap effect
should be reflected in the low-energy electronic structure. Here the effect of the
normal-state pseudogap on the low-energy electronic structure in the underdoped
cuprate superconductors is investigated within the framework of the
kinetic-energy-driven superconductivity. The strong coupling of the electrons with
the spin excitation induces the normal-state pseudogap-state in the particle-hole
channel and superconducting (SC) state in the particle-particle channel, where the
normal-state pseudogap and SC gap respectively originate from the electron normal
and anomalous self-energies, and are evaluated by taking into account the vertex
correction. As a natural consequence of the interplay between the normal-state
pseudogap-state and SC-state, the SC transition temperature $T_{\rm c}$ exhibits a
dome-like shape of the doping dependence, however, in a striking contrast to
$T_{\rm c}$ in the underdoped regime, the normal-state pseudogap crossover
temperature $T^{*}$ is much higher than $T_{\rm c}$ in the underdoped regime, and
then it decreases with the increase of doping, eventually disappearing together with
$T_{\rm c}$ at the end of the SC dome. Concomitantly, the spectral weight on the
electron Fermi surface (EFS) at around the antinodal region is suppressed strongly
by this normal-state pseudogap, and then EFS is truncated to form four disconnected
Fermi arcs centered around the nodal region with the largest spectral weight located
at around the tips of the disconnected Fermi arcs. Moreover, the dip in the
peak-dip-hump structure observed in the energy distribution curve and checkerboard
charge ordering found in the ARPES autocorrelation are intrinsically connected with
the emergence of the normal-state pseudogap. The theory therefore indicates that the
same spin excitation that governs both the normal-state pseudogap-state and SC-state
naturally leads to the exotic features of the low-energy electronic structure in the
underdoped cuprate superconductors.
\end{abstract}


\maketitle

\section{Introduction}\label{Introduction}

In the conventional superconductors\cite{Bardeen57,Schrieffer64}, only an energy gap
emerges in the quasiparticle excitation spectrum below the superconducting (SC)
transition temperature $T_{\rm c}$. This energy gap is corresponding to the energy to
break an electron pair, and therefore is identified as the electron-pair gap or the SC
gap. As in the case of the conventional superconductors\cite{Bardeen57,Schrieffer64},
the pairing of electrons in the cuprate superconductors occurs at $T_{\rm c}$,
creating the SC gap that serves as the SC-state order parameter
\cite{Bednorz86,Wu87,Tsuei00}. However, in a stark contrast to the conventional
superconductors\cite{Bardeen57,Schrieffer64}, a distinct energy gap called the
normal-state pseudogap
\cite{Puchkov96,Timusk99,Hufner08,Hussey08,Damascelli03,Campuzano04,Fink07,Hussey11,Barisic13,Keimer15,Vishik18}
exists above $T_{\rm c}$ but below the normal-state pseudogap crossover temperature
$T^{*}$. In particular, this normal-state pseudogap is most notorious in the
underdoped regime,
where the charge-carrier doping concentration is too low for the optimal
superconductivity. This is why in the underdoped regime, the phase above $T_{\rm c}$
but below $T^{*}$ is so-called as {\it the normal-state pseudogap phase}. This
normal-state pseudogap is the most noteworthy, which suppresses strongly the
electronic density of states on the electron Fermi surface (EFS), and then the physical
response in the underdoped regime can be well interpreted in terms of the formation of
this normal-state pseudogap.

By virtue of systematic studies using angle-resolved photoemission spectroscopy
(ARPES) measurement techniques, a number of consequences from the effect of the
normal-state pseudogap on the low-energy electronic structure together with the
associated exotic phenomena have been identified experimentally in the underdoped
regime, where the main characteristic
features can be summarized as: (i) the stronger quasiparticle scattering is observed
at around the antinodal region than at around the nodal region, this leads to the
formation of the disconnected Fermi arcs centered at around the nodal region
\cite{Norman98,Yoshida06,Tanaka06,Kanigel07,Shi08,Sassa11,Horio16,Loret18,Comin14}.
In particular, one of the intrinsic features associated with this EFS instability is
the emergence of the multiple nearly-degenerate electronic orders
\cite{Comin14,Gh12,Neto14,Hash15,Comin16,Campi15}; (ii) the dramatic change in the
line-shape of the energy distribution curve is detected
\cite{Dessau91,Campuzano99,Lu01,Sato02,Borisenko03,Wei08,Sakai13,Hashimoto15,Loret17,DMou17},
where a sharp quasiparticle peak develops at low binding energies, followed by a
dip and a hump, giving rise to the so-called peak-dip-hump (PDH) structure; (iii)
the sharp peaks in the ARPES autocorrelation spectrum in the SC-state with the
scattering wave vectors ${\bf q}_{i}$ are directly correlated with the regions of
the highest joint density of states\cite{Chatterjee06,He14,Restrepo23}, where
scattering wave vectors ${\bf q}_{i}$ connect the tips of the Fermi arcs obeying an
{\it octet scattering model}. Since the anomalous properties, including the
exceptionally strong superconductivity, have often been attributed to the
particular characteristics of the low-energy electronic structure
\cite{Damascelli03,Campuzano04,Fink07}, the understanding of the effect of the
normal-state pseudogap on the low-energy electronic structure in the underdoped
cuprate superconductors is thought to be key to the understanding of the
nonconventional mechanism of superconductivity.

Although the exotic features of the low-energy electronic structure in the
underdoped cuprate superconductors have been established experimentally, a complete
understanding of these exotic features is still unclear. After intensive
investigations over three decades, a large body of experimental evidence has shown
that as in the case of the SC-state\cite{Carbotte11,Bok16}, the normal-state
pseudogap-state may also originate from the electron interaction mediated by a
particular bosonic mode
\cite{Puchkov96,Timusk99,Hufner08,Hussey08,Damascelli03,Campuzano04,Fink07,Hussey11,Barisic13,Keimer15,Vishik18}.
In this case, a key question is whether a common bosonic excitation, which mediates
the interaction between electrons responsible for both the SC-state and normal-state
pseudogap-state, generates the anomalous properties of the electronic structure. In
the previous works\cite{Feng16,Gao18,Gao19}, the low-energy electronic structure in
the cuprate superconductors has been discussed within the framework of the
kinetic-energy-driven superconductivity, where both the electron normal self-energy
in the particle-hole channel and electron anomalous self-energy in the
particle-particle channel generated by the coupling of the electrons with the
spin excitation are evaluated, and employed to study the renormalization of the
electrons. However, these calculations\cite{Feng16,Gao18,Gao19} suffer from ignoring
the vertex correction for the electron normal and anomalous self-energies, while
the influence of the vertex correction may be dramatic in the underdoped regime.
In this paper, we study the low-energy electronic structure in the underdoped cuprate
superconductors along with this line\cite{Feng16,Gao18,Gao19} by taking into account
the vertex correction for the electron normal and anomalous self-energies, where we
show clearly that (i) the normal-state pseudogap and SC gap respectively originate
from the electron normal self-energy in the particle-hole channel and anomalous
self-energy in the particle-particle channel, and then the SC-state coexists
and competes with the normal-state pseudogap-state below $T_{\rm c}$ over the whole
SC dome. This leads to that although $T_{\rm c}$ takes a dome-like shape with the
underdoped and the overdoped regimes on each side of the optimal doping, where
$T_{\rm c}$ reaches its maximum, $T^{*}$ is much higher than $T_{\rm c}$ in
the underdoped regime, and then it gradually merges with the SC gap in the strongly
overdoped region; (ii) the characteristic features of the low-energy electronic
structure observed from the ARPES experiments, including the EFS reconstruction
\cite{Norman98,Yoshida06,Tanaka06,Kanigel07,Shi08,Sassa11,Horio16,Loret18,Comin14},
the nature of the charge-order correlation
\cite{Comin14,Gh12,Neto14,Hash15,Comin16,Campi15},
the striking PDH structure in the energy distribution curve
\cite{Dessau91,Campuzano99,Lu01,Sato02,Borisenko03,Wei08,Sakai13,Hashimoto15,Loret17,DMou17},
and the remarkable ARPES autocorrelation spectrum\cite{Chatterjee06,He14,Restrepo23},
are qualitatively reproduced. In particular, in addition to these sharp peaks in the
ARPES autocorrelation spectrum with the scattering wave vectors ${\bf q}_{i}$, the
notable checkerboard peaks appear only in the normal-state pseudogap phase, and are
directly linked with the opening of the normal-state pseudogap. Our results therefore
also indicate that the {\it same spin excitation} that governs both the normal-state
pseudogap-state and the SC-state naturally generates the exotic features of the
low-energy electronic structure in the underdoped cuprate superconductors.

This paper is organized as follows. The theoretical framework is presented in Section
\ref{framework}, where the full electron diagonal and off-diagonal propagators
obtained by taking into account the vertex correction are given. Starting from these
full electron diagonal and off-diagonal propagators, the phase diagram of the cuprate
superconductors is derived for a convenience in the discussion of the low-energy
electronic structure in the underdoped cuprate superconductors. The quantitative
characteristics of the low-energy electronic structure are presented in Section
\ref{Electronic-structure}, where we show that the scattering wave vector of the
notable checkerboard charge ordering found in the ARPES autocorrelation in the
normal-state pseudogap phase is well consistent with that observed from the ARPES and
scanning tunneling spectroscopy (STS) experiments. Finally, we give a summary in
Sec. \ref{Summary}. In Appendix
\ref{Derivation-of-propagators}, we present the generalization of the main
formalisms of the kinetic-energy-driven superconductivity from the case with the
neglect of the vertex correction for the electron normal and anomalous self-energies
to the present case by the inclusion of the vertex correction, where the detail of
the calculation of the phase diagram is given.

\section{Methodology}\label{framework}

\subsection{Model and electron local constraint}\label{model-constraint}

The basic element of the crystal structure in cuprate superconductors is the
two-dimensional CuO$_{2}$ plane\cite{Bednorz86,Wu87} in which superconductivity with
the exceptionally high $T_{\rm c}$ appears upon charge-carrier doping. Soon after the
discovery of superconductivity in cuprate superconductors,
Anderson\cite{Anderson87} recognized that the essential physics of cuprate
superconductors can be properly depicted in terms of the $t$-$J$ model on a square
lattice,
\begin{equation}\label{tjham}
H=-\sum_{ll'\sigma}t_{ll'}C^{\dagger}_{l\sigma}C_{l'\sigma}
+\mu\sum_{l\sigma}C^{\dagger}_{l\sigma}
C_{l\sigma}+J\sum_{\langle ll'\rangle}{\bf S}_{l}\cdot {\bf S}_{l'},~~
\end{equation}
where $C^{\dagger}_{l\sigma}$ ($C_{l\sigma}$) creates (annihilates) an electron with
spin index $\sigma$ on lattice site $l$, ${\bf S}_{l}$ is a localized spin operator
with its components $S_{l}^{x}$, $S_{l}^{y}$, and $S_{l}^{z}$, and $\mu$ is the
chemical potential. The kinetic-energy term consists of the constrained electron
hopping between the nearest-neighbor (NN) sites $\hat{\eta}$ with the hoping integral
$t_{ll'}=t_{\hat{\eta}}=t$ and the constrained electron hopping between the next NN
sites $\hat{\tau}$ with the hoping integral $t_{ll'}=t_{\hat{\tau}}=-t'$, while the
magnetic-energy term consists of the magnetic interaction between the NN sites
$\hat{\eta}$ with the magnetic exchange coupling $J$. The summation $ll'$ indicates
that $l$ runs over all sites, and for each $l$, over its NN sites $\hat{\eta}$ and
next NN sites $\hat{\tau}$, while the summation $\langle ll'\rangle$ is taken over all
the NN pairs. In the following discussions, the NN magnetic exchange coupling $J$ and
the lattice constant of the square lattice are respectively set as the energy and
length units, while $t$ and $t'$ are respectively set to be $t/J=2.5$ and $t'/t=0.3$,
which are the typical values of cuprate superconductors
\cite{Damascelli03,Campuzano04,Fink07}. In particular, when necessary to compare with
the experimental data, we set $J=100$meV.

The $t$-$J$ model (\ref{tjham}) is the strong-coupling limit of the Hubbard model,
and then the difficulty of its solution lies in enforcing the no double electron
occupancy on-site local constraint\cite{Yu92,Feng93,Zhang93,Lee06,Spalek22}, i.e.,
$\sum_{\sigma}C^{\dagger}_{l\sigma}C_{l\sigma}\leq 1$. An intuitively appealing
approach to implement this on-site local constraint is the slave-particle
approach\cite{Yu92,Feng93,Zhang93,Lee06}, where the constrained electron operator is
decoupled as $C_{l\sigma}$=$a^{\dagger}_{l}f_{l\sigma}$ with $a^{\dagger}_{l}$ as
the slave boson and $f_{l\sigma}$ as the spinful fermion and the on-site local
constraint $\sum_{\sigma}f^{\dagger}_{l\sigma}f_{l\sigma}+a^{\dagger}_{l}a_{l}=1$,
or {\it vice versa}, {\it i.e.}, $a^{\dagger}_{l}$ as the fermion and $f_{l\sigma}$
as the boson. Due to the on-site local constraint, these particles are also coupled
by a strong $U(1)$ gauge field\cite{Yu92,Lee06}, allowed in this slave-particle
representation. However, there are a number of difficulties in this approach. First
of all, in the slave-boson version, the antiferromagnetic correlation is absent for
zero doping, so the ground-state energy is high compared with the numerical
estimation for small clusters, and the Marshall sign-rule is not obeyed.
Alternatively, in the slave-fermion approach, the ground-state is antiferromagnetic
for the undoped case and persists until very high doping ($\sim 60\%$). In
particular, if we, following the common practice, let $f_{l\sigma}$ keep track of
the spin degree of freedom of the constrained electron, while $a_{l}$ keep track of
the charge degree of freedom of the constrained electron, satisfying the sum rules:
$\delta = <a^{\dagger}_{l}a_{l}>$ and
$1-\delta = \sum_{\sigma}<f^{\dagger}_{l\sigma}f_{l\sigma}>$, where $\delta$ is the
charge-carrier doping concentration, we\cite{Feng93} find that the sum rule for the
physical electron $\sum_{\sigma}<C^{\dagger}_{l\sigma}C_{l\sigma}>=1-\delta$ is not
satisfied for both versions. Moreover, we\cite{Feng93} have also shown that the
overall electron distribution does not have the appropriate EFS within this scheme.
These are intrinsic difficulties of this decoupling scheme. To avoid these
difficulties in the slave-particle approach, the method employed to analyse the
$t$-$J$ model (\ref{tjham}) together with the on-site local constraint of the no
double electron occupancy in this paper is the fermion-spin
theory\cite{Feng9404,Feng15}, in which the constrained electron operators
$C_{l\uparrow}$ and $C_{l\downarrow}$ in the $t$-$J$ model (\ref{tjham}) are
decoupled into two distinct operators as,
\begin{eqnarray}\label{CSS}
C_{l\uparrow}=h^{\dagger}_{l\uparrow}S^{-}_{l}, ~~~~
C_{l\downarrow}=h^{\dagger}_{l\downarrow}S^{+}_{l},
\end{eqnarray}
where $h^{\dagger}_{l\sigma}=e^{i\Phi_{l\sigma}}h^{\dagger}_{l}$
($h_{l\sigma}=e^{-i\Phi_{l\sigma}}h_{l}$) is a spinful fermion creation (annihilation)
operator, which creates (annihilates) a charge carrier at site $l$, and therefore
keeps track of the charge degree of freedom of the constrained electron together with
some effects of spin configuration rearrangements due to the presence of the doped
charge carrier itself, while $S^{+}_{l}$ ($S^{-}_{l}$) is the spin-raising
(spin-lowering) operator, which keeps track of the spin degree of freedom of the
constrained electron. The advantages of this fermion-spin approach (\ref{CSS}) can be
summarized as: (i) the on-site local constraint of no double occupancy is satisfied
in actual analyses; (ii) the charge carrier or spin {\it itself} is $U(1)$ gauge
invariant, and then the collective mode for the spin is real and can be interpreted
as the spin excitation responsible for the dynamical spin response, while the electron
quasiparticle as a result of the charge-spin recombination of a charge carrier and a
localized spin is not affected by the statistical $U(1)$ gauge fluctuation, and is
responsible for the electronic-state properties. This is why the fermion-spin approach
(\ref{CSS}) is an efficient calculation scheme which can provide a proper description
of the anomalous properties of cuprate superconductors.

\subsection{Interplay between superconducting-state and normal-state pseudogap-state}
\label{IBSPS}

In a superconductor, the formation of the electron pairs is crucial in the occurrence
of superconductivity because the electron pairs behave as effective bosons, and can
form something analogous to a Bose condensate that flows without resistance
\cite{Anderson07}. This follows from a basic fact that although electrons repel each
other because of the presence of the Coulomb repulsive interaction, at low energies
and low temperatures, there can be an effective attraction originating due to the
exchange of boson excitations\cite{Anderson07}. In the conventional superconductors,
as explained by the Bardeen-Cooper-Schrieffer (BCS) theory\cite{Schrieffer64,Bardeen57},
these exchanged boson excitations are {\it phonons}. These phonons act like a
{\it bosonic glue} to hold the electron pairs together\cite{Cooper56}, and then the
condensation of
these electron pairs reveals the SC-state\cite{Schrieffer64,Bardeen57}. However, the
BCS theory is not specific to a phonon-mediated interaction between electrons, other
bosonic excitations can also serve as the pairing glue\cite{Miller11,Monthoux07}. In the
past over three decades, a series of experiments from the nuclear magnetic resonance,
nuclear quadrupole resonance, inelastic neutron scattering, and resonant inelastic
X-ray scattering measurements on the cuprate superconductors
\cite{Fujita12,Birgeneau89,Fong95,Yamada98,Arai99,Bourges00,He01,Tranquada04,Bourges05,Dean15}
have identified the spin excitations with the high intensity over a large part of
momentum space. More importantly, these spin excitations exist across the entire range of
the SC dome, and with sufficient intensity to mediate superconductivity. In this case,
a crucial question is whether the spin excitation can mediate electron pairing in the
cuprate superconductors in analogy to the phonon-mediated pairing mechanism in the
conventional superconductors? Starting from the fermion-spin theory (\ref{CSS})
description of the $t$-$J$ model (\ref{tjham}), we\cite{Feng15,Feng0306,Feng12,Feng15a}
have established a kinetic-energy-driven SC mechanism, where the d-wave charge-carrier
pairs are generated by the interaction between charge carriers directly from the
kinetic energy of the $t$-$J$ model by the exchange of the spin excitation, then the
d-wave electron pairs originate from the d-wave charge-carrier pairs due to
charge-spin recombination\cite{Feng15a}, and these electron pairs condense to form
the d-wave SC-state. The mechanism of the kinetic-energy-driven superconductivity
is purely electronic without phonon, since the main ingredient is identified into an
electron pairing mechanism not involving the phonon, the external degree of freedom,
but {\it the spin excitation, the collective mode from the internal spin degree of
freedom of the constrained electron itself}. In other words, the constrained electrons
simultaneously act to glue and to be glued\cite{Schrieffer95,Xu23a}. Moreover, this
kinetic-energy-driven SC-state in a way is in turn strongly influenced by the
single-particle coherence, which leads to that the maximal $T_{\rm c}$ occurs at
around the optimal doping, and then decreases in both the underdoped and overdoped
regimes. Our following discussions of the interplay between the SC-state and the
normal-state pseudogap-state in the cuprate superconductors builds on the
kinetic-energy-driven superconductivity, where the electron normal and anomalous
self-energies in this paper are derived by taking into account
the vertex correction. For the following discussions of the main physics in the core,
the details of the derivation of the main formalisms of the kinetic-energy-driven
superconductivity and the related vertex-corrected electron normal and anomalous
self-energies are presented in Appendix \ref{Derivation-of-propagators}. Following
our previous discussions\cite{Feng15a}, the full electron diagonal and off-diagonal
propagators of the $t$-$J$ model (\ref{tjham}) in the fermion-spin representation
(\ref{CSS}) can be derived explicitly as
[see Appendix \ref{Derivation-of-propagators}],
\begin{subequations}\label{EDODGF}
\begin{eqnarray}
G({\bf k},\omega)&=&{1\over\omega-\varepsilon_{\bf k}
-\Sigma_{\rm tot}({\bf k},\omega)},~~~~~\label{EDGF}
\end{eqnarray}
\begin{eqnarray}
\Im^{\dagger}({\bf k},\omega)&=&{W({\bf k},\omega)\over\omega
-\varepsilon_{\bf k}-\Sigma_{\rm tot}({\bf k},\omega)},
~~~~~~~\label{EODGF}
\end{eqnarray}
\end{subequations}
where $\varepsilon_{\bf k}=-4t\gamma_{\bf k}+4t'\gamma_{\bf k}'+\mu$ is the electron
energy dispersion in the tight-binding approximation, with
$\gamma_{\bf k}=({\rm cos}k_{x}+{\rm cos} k_{y})/2$, and
$\gamma_{\bf k}'={\rm cos}k_{x}{\rm cos}k_{y}$, while the electron total self-energy
$\Sigma_{\rm tot}({\bf k},\omega)$ and the function $W({\bf k},\omega)$ are given
explicitly as,
\begin{subequations}\label{HTOTSE}
\begin{eqnarray}
\Sigma_{\rm tot}({\bf k},\omega)&=& \Sigma_{\rm ph}({\bf k},\omega)
+{|\Sigma_{\rm pp}({\bf k},\omega)|^{2}\over\omega +\varepsilon_{\bf k}
+\Sigma_{\rm ph}({\bf k},-\omega)},~~~~~\\
W({\bf k},\omega)&=&-{\Sigma_{\rm pp}({\bf k},\omega)\over\omega+\varepsilon_{\bf k}
+\Sigma_{\rm ph}({\bf k},-\omega)},
\end{eqnarray}
\end{subequations}
respectively, with the electron normal self-energy $\Sigma_{\rm ph}({\bf k},\omega)$
in the particle-hole channel and electron anomalous self-energy
$\Sigma_{\rm pp}({\bf k},\omega)$ in the particle-particle channel. In the previous
derivations\cite{Feng16,Gao18,Gao19}, the electron normal and anomalous self-energies
have been calculated with neglecting the vertex correction, where the obtained
results of the electronic structure in the SC-state is qualitatively consistent with
the corresponding experimental results observed in the ARPES measurements
\cite{Dessau91,Campuzano99,Lu01,Sato02,Borisenko03,Wei08,Sakai13,Hashimoto15,Loret17,DMou17,Chatterjee06,He14,Restrepo23},
however, the magnitude of the pseudogap gap in the underdoped regime is not large
enough compared with the corresponding experimental observations\cite{Vishik18},
leading to that the electronic density of states on EFS in the underdoped regime is
not sufficiently suppressed. This weakness can be corrected in terms of the vertex
correction of the electron self-energy\cite{Zeng25}. In this paper, for a more
proper description of the interplay between the SC-state and normal-state
pseudogap-state and the related effect of the normal-state pseudogap on the low-energy
electronic structure in the
underdoped cuprate superconductors, the above electron normal self-energy
$\Sigma_{\rm ph}({\bf k},\omega)$ and electron anomalous self-energy
$\Sigma_{\rm pp}({\bf k},\omega)$ in Eq. (\ref{HTOTSE}) can be respectively derived
by taking into account the vertex corrections as\cite{Zeng25}
[see Appendix \ref{Derivation-of-propagators}],
\begin{subequations}\label{E-N-AN-SE}
\begin{eqnarray}
\Sigma_{\rm ph}({\bf k},i\omega_{n})&=&{1\over N^{2}}\sum_{{\bf p},{\bf p}'}
[V_{\rm ph}\Lambda_{{\bf p}+{\bf p}'+{\bf k}}]^{2}\nonumber\\
&\times& {1\over\beta}\sum_{ip_{m}}G({{\bf p}+{\bf k}},ip_{m}+i\omega_{n})
\Pi({\bf p},{\bf p}',ip_{m}),\nonumber\\
~~~~~~~~~~\label{E-N-SE}\\
\Sigma_{\rm pp}({\bf k},i\omega_{n})&=&{1\over N^{2}}\sum_{{\bf p},{\bf p}'}
[V_{\rm pp}\Lambda_{{\bf p}+{\bf p}'+{\bf k}}]^{2}\nonumber\\
&\times& {1\over \beta}\sum_{ip_{m}}\Im^{\dagger}({\bf p}+{\bf k},ip_{m}+i\omega_{n})
\Pi({\bf p},{\bf p}',ip_{m}),\nonumber\\
~~~~~~~~~~\label{E-AN-SE}
\end{eqnarray}
\end{subequations}
where $\Lambda_{\bf k}=4t\gamma_{\bf k}-4t'\gamma_{\bf{k}}'$ is the bare vertex
function, while $V_{\rm ph}$ is the vertex correction for the electron normal
self-energy, and $V_{\rm pp}$ is the vertex correction for the electron anomalous
self-energy, which together with the spin bubble $\Pi({\bf p},{\bf p}',ip_{m})$ have
been determined explicitly in Appendix \ref{Derivation-of-propagators}
[see Appendix \ref{Derivation-of-propagators}]. Moreover,
the sharp peaks appear at low-temperature in $\Sigma_{\rm ph}({\bf k},\omega)$,
$\Sigma_{\rm pp}({\bf k},\omega)$, and the related quantities that are actually a
$\delta$-function, which are broadened by a small damping employed in the numerical
calculation for a finite lattice\cite{Brinckmann01}. The calculation in this paper
for $\Sigma_{\rm ph}(\bf k,\omega)$, $\Sigma_{\rm pp}({\bf k},\omega)$, and the
related quantities is performed numerically on a $120\times 120$ lattice in momentum
space, where the infinitesimal $i0_{+}\rightarrow i\Gamma$ is replaced by a small
damping $\Gamma=0.05J$.

Although both the electron normal and anomalous self-energies originate from the
coupling of the electrons with the same spin excitation, they capture different
aspects of the interaction effect. On the one hand, the SC-state is characterized
by the electron anomalous self-energy $\Sigma_{\rm pp}({\bf k},\omega)$, which is
identified as the energy and momentum dependent electron pair gap (then the SC gap)
$\Sigma_{\rm pp}({\bf k},\omega)=\bar{\Delta}_{\bf k}(\omega)$, and thus is
corresponding to the energy for breaking an electron pair\cite{Cooper56}. In the
static limit, this electron pair gap $\bar{\Delta}_{\bf k}(\omega)$ is reduced as
$\bar{\Delta}_{\bf k}(\omega)\mid_{\omega=0}=\bar{\Delta}_{\bf k}=\bar{\Delta}
\gamma^{\rm (d)}_{\bf k}$, with the electron pair gap parameter $\bar{\Delta}$ and
the d-wave factor $\gamma^{\rm (d)}_{\bf k}=({\rm cos}k_{x}-{\rm cos} k_{y})/2$
[see Appendix \ref{Derivation-of-propagators}]. On the other hand, the single-particle
coherence is dominated by the electron normal self-energy
$\Sigma_{\rm ph}({\bf k},\omega)$, which gives rise to a main contribution to the
energy and lifetime renormalization of the electrons, and then all the anomalous
properties of the cuprate superconductors arise from this renormalization of the
electrons.

Among the unusual features that are unique to the cuprate superconductors, in addition
to the exceptionally strong superconductivity, the normal-state pseudogap is
perhaps the most noteworthy, however, the origin of the normal-state pseudogap
remains a subject of debate. During the last over three decades, several scenarios
were proposed to explain the formation of the normal-state pseudogap. In particular,
it has been argued\cite{Emery95} that the normal-state pseudogap originates
from preformed pairs at $T^{*}$, which would then condense (that is, become phase
coherent) at $T_{\rm c}$. On the other hand, it has been suggested that the
normal-state pseudogap is distinct from the SC gap and related with a certain order
which competes with superconductivity\cite{Kondo09}. Moreover, it has been proposed
that the normal-state pseudogap is a combination of a quantum disordered d-wave
superconductor and an entirely different form of competing order, originating from
the particle-hole channel\cite{Tesanovic08}. In the present framework of the
kinetic-energy-driven superconductivity, the normal-state pseudogap is intimately
connected to the electron normal self-energy. To see this intimate connection more
clearly, the above electron normal self-energy in Eq. (\ref{E-N-SE}) can be also
rewritten as,
\begin{equation}\label{ESE-NSPG}
\Sigma_{\rm ph}({\bf k},\omega) \approx {[2\bar{\Delta}_{\rm PG}({\bf k})]^{2}
\over\omega -\varepsilon_{\bf k}},
\end{equation}
and then the momentum-dependent normal-state pseudogap
$\bar{\Delta}_{\rm PG}({\bf k})$ is obtained straightforwardly as,
\begin{equation}\label{NSPG}
\bar{\Delta}^{2}_{\rm PG}({\bf k})=-{1\over 4}\varepsilon_{\bf k}
\Sigma_{\rm ph}({\bf k},0),
\end{equation}
with the normal-state pseudogap parameter
$\bar{\Delta}^{2}_{\rm PG}=(1/N)\sum_{\bf k}\bar{\Delta}^{2}_{\rm PG}({\bf k})$.
This momentum-dependent normal-state pseudogap is identified as being a region of
the electron normal self-energy in which the normal-state pseudogap anisotropically
suppresses the electronic density of states on EFS. In particular, the present
results also show that the same interaction between electrons mediated by the spin
excitation that is responsible for pairing the electrons in the particle-particle
channel also generates the normal-state pseudogap state in the particle-hole
channel\cite{Feng12,Feng15a}. Moreover, it should be noted that in spite of the
origins of the formation of the normal-state pseudogap, the main feature of the full
electron propagators in Eq. (\ref{EDODGF}) together with the SC gap
$\bar{\Delta}_{\bf k}$ and the electron normal self-energy (then the normal-state
pseudogap) in Eq. (\ref{ESE-NSPG}) are similar to these proposed from several
phenomenological theories\cite{Benfatto00,Cho06,Millis06,Rice12,Tesanovic08} of
the normal-state pseudogap-state based on the d-wave BCS formalism, where a
phenomenological normal-state pseudogap is introduced to describe the two-gap feature
in cuprate superconductors.

\begin{figure}[h!]
\includegraphics[scale=0.32]{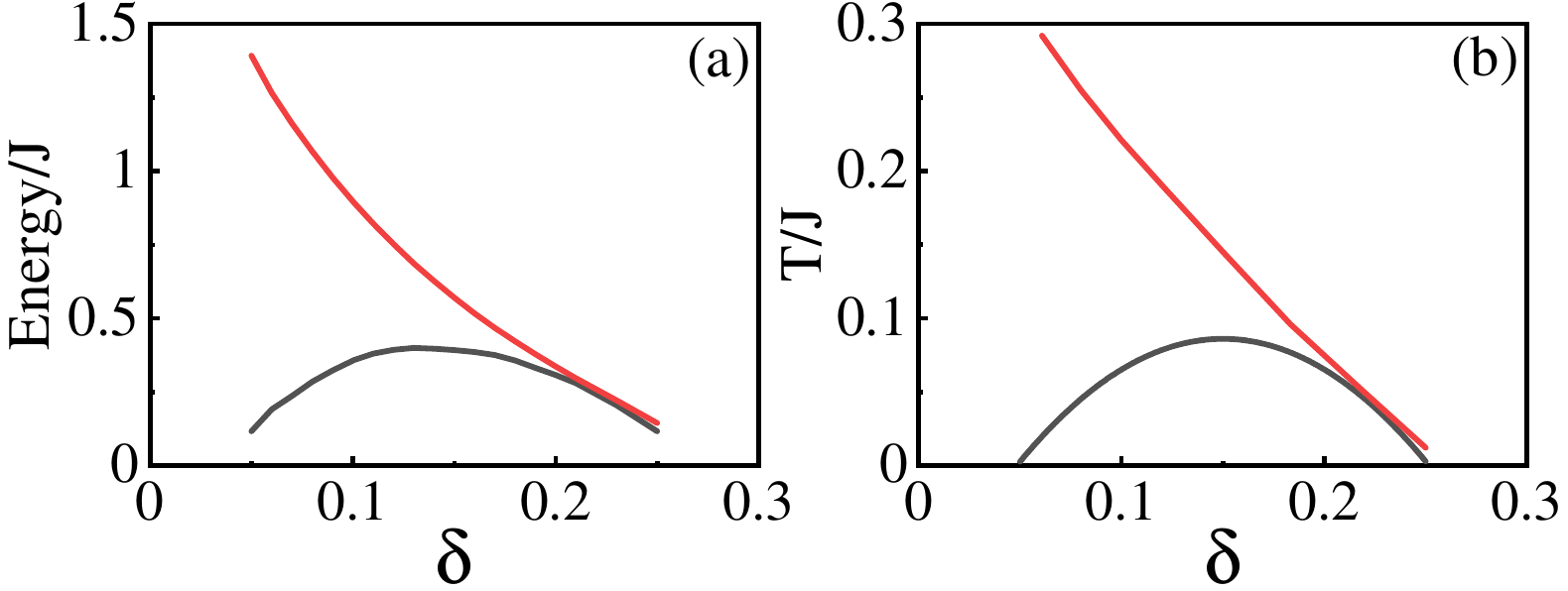}
\caption{(Color online) (a) The electron pair gap $2\bar{\Delta}$ (black-line) and
normal-state pseudogap $2\bar{\Delta}_{\rm PG}$ (red-line) as a function of doping
with temperature $T=0.002J$. (b) The superconducting transition temperature
$T_{\rm c}$ (black-line) and the normal-state pseudogap crossover temperature $T^{*}$
(red-line) as a function of doping.\label{EPG-T-doping}}
\end{figure}

For a better understanding of the interplay between the SC-state and normal-state
pseudogap-state in the cuprate superconductors, we have performed a series of
calculations for the electron pair gap parameter $\bar{\Delta}$ and normal-state
pseudogap parameter $\bar{\Delta}_{\rm PG}$ at different doping levels [see Appendix
\ref{Derivation-of-propagators}], and the results of 2$\bar{\Delta}$ (black-line) and
2$\bar{\Delta}_{\rm PG}$ (red-line) as a function of doping with temperature
$T=0.002J$ are plotted in Fig. \ref{EPG-T-doping}a. Apparently, the exotic two-gap
feature detected from the cuprate superconductors\cite{Hufner08}
is qualitatively reproduced, where with the increase of doping, the electron pair gap
$\bar{\Delta}$ is raised monotonically in the underdoped regime, and reaches its
maximum at around the optimal doping $\delta\approx 0.15$. However, with the further
increase of doping, $\bar{\Delta}$ then turns into a gradual decrease in the overdoped
regime. On the other hand, in contrast to the case of $\bar{\Delta}$ in the underdoped
regime, the normal-state pseudogap $\bar{\Delta}_{\rm PG}$ in the underdoped regime is
much larger than $\bar{\Delta}$ in the underdoped regime, reflecting a basic
experimental fact that the normal-state pseudogap-state is particularly obvious in the
underdoped regime. This normal-state pseudogap $\bar{\Delta}_{\rm PG}$ smoothly
decreases with the increase of doping. In particular, $\bar{\Delta}_{\rm PG}$ seems to
merge with $\bar{\Delta}$ in the strongly overdoped region, eventually going to zero
together with $\bar{\Delta}$ at the end of the SC dome.

Both the above electron pair gap and normal-state pseudogap are strong temperature
dependent. In particular, in a given doping concentration, the electron pair gap
vanishes at $T_{\rm c}$, while the normal-state pseudogap disappears when temperature
reaches the normal-state pseudogap crossover temperature $T^{*}$. To see the
different doping-dependent trends of $T_{\rm c}$ and $T^{*}$ in the underdoped
regime more clearly, we have also made a series of calculations for $T_{\rm c}$ and
$T^{*}$ at different doping concentrations [see Appendix
\ref{Derivation-of-propagators}], and the results of
$T_{\rm c}$ (black-line) and $T^{*}$ (red-line) as a function of doping are plotted
in Fig. \ref{EPG-T-doping}b, where in conformity with the doping dependence of
$\bar{\Delta}$ and $\bar{\Delta}_{\rm PG}$ shown in Fig. \ref{EPG-T-doping}a, (i)
$T_{\rm c}$ increases with the increase of doping in the underdoped regime, and
reaches its maximum at around the optimal doping, then decreases in the overdoped
regime\cite{Feng15a}; (ii) $T^{*}$ is much larger than $T_{\rm c}$ in the underdoped
regime, then it smoothly decreases with the increase of doping, eventually
terminating together with $T_{\rm c}$ at the end of the SC dome, in qualitative
agreement with the experimental observations\cite{Hufner08}.
This normal-state pseudogap crossover temperature $T^{*}$ is actually a crossover
line below which a novel electronic state emerges, as exemplified by the presence of
the disconnected Fermi arcs, the competing electronic orders, the PDH structure in
the energy distribution curve, etc., and then the anomalous properties
correlated to the formation of the normal-state pseudogap discussed in the next
Sec. \ref{Electronic-structure} are explained in a natural way.

\section{Quantitative characteristics of low-energy electronic structure}
\label{Electronic-structure}

In this section, we discuss the quantitative characteristics of the low-energy
electronic structure in the underdoped cuprate superconductors, and show how the
normal-state pseudogap leads to the crucial effect on the redistribution of the
spectral weight on EFS, the energy distribution curves, and the ARPES
autocorrelation.

The anomalous properties of the electronic structure observed from ARPES experiments
can be analyzed theoretically in terms of the ARPES spectral
intensity\cite{Damascelli03,Campuzano04,Fink07},
\begin{equation}\label{QPES}
I({\bf k},\omega)\propto n_{\rm F}(\omega) A({\bf k},\omega),
\end{equation}
with the fermion distribution $n_{\rm F}(\omega)$ and the electron spectral
function,
\begin{equation}\label{ESF}
A({\bf k},\omega)={1\over\pi}{\tilde{\Gamma}_{\bf k}(\omega)\over
[\omega-\varpi_{\bf k}(\omega)]^{2}+[\tilde{\Gamma}_{\bf k}(\omega)]^{2}},~~~
\end{equation}
where the quasiparticle energy dispersion $\varpi_{\bf k}(\omega)$ and quasiparticle
scattering rate $\tilde{\Gamma}_{\bf k}(\omega)$ in the SC-state can be
expressed explicitly as,
\begin{subequations}\label{QPED-QPSR}
\begin{eqnarray}
\varpi_{\bf k}(\omega)&=&\varepsilon_{\bf k}
-{\rm Re}\Sigma_{\rm tot}({\bf k},\omega),\label{QPED}\\
\tilde{\Gamma}_{\bf k}(\omega)&=& {\rm Im}\Sigma_{\rm tot}({\bf k},\omega),
~~~~~~\label{QPSR}
\end{eqnarray}
\end{subequations}
with ${\rm Re}\Sigma_{\rm tot}({\bf k},\omega)$ and
${\rm Im}\Sigma_{\rm tot}({\bf k},\omega)$ that are respectively the real and
imaginary parts of the electron total self-energy $\Sigma_{\rm tot}({\bf k},\omega)$.
The above results in Eq. (\ref{QPED-QPSR}) show that the constrained electrons in the
cuprate superconductors are renormalized to form the quasiparticles due to the
electron scattering mediated by the spin excitation and therefore they acquire a
finite lifetime.

\subsection{Electron Fermi surface reconstruction}\label{EFSR-NS}

\begin{figure}[h!]
\includegraphics[scale=0.25]{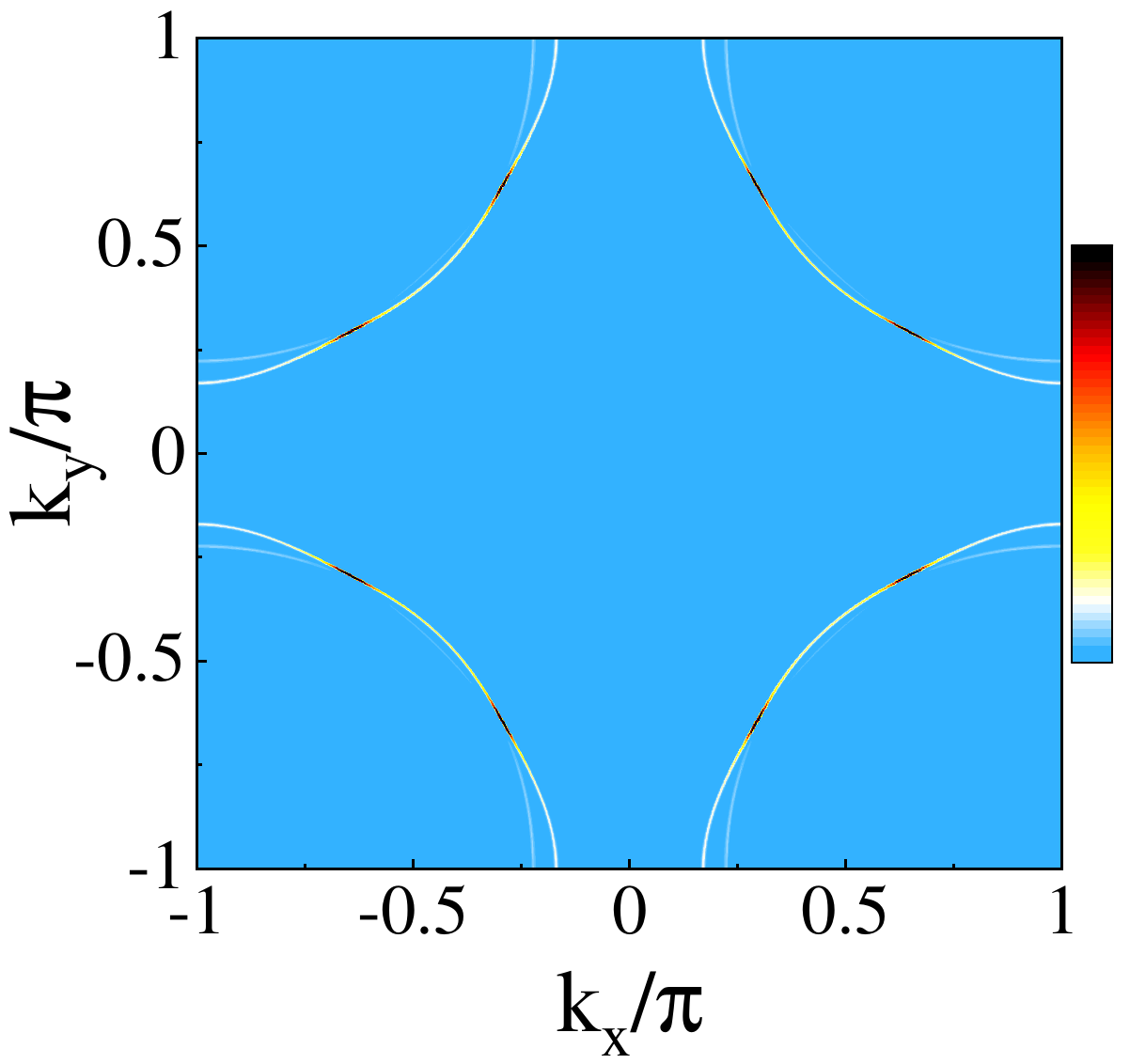}
\caption{(Color online) The intensity map of the normal-state ARPES spectrum in
$[k_{x},k_{y}]$ plane for the zero binding-energy $\omega=0$ at $\delta=0.10$ with
$T=0.002J$.  \label{EFS-maps}}
\end{figure}

EFS is usually defined in the normal-state, which separates the filled electronic
states from the empty electronic states, and therefore plays a key role in the
understanding of the anomalous properties in the cuprate superconductors.
Experimentally, the {\it intensity map} of the ARPES spectrum at the zero energy is
often adopted to measure the underlying EFS\cite{Damascelli03,Campuzano04,Fink07},
i.e., the underlying EFS is determined by measuring the ARPES spectrum at the zero
energy to map out the locus of the maximum in the intensity.

In the normal-state ($\bar{\Delta}=0$), the quasiparticle energy dispersion
and quasiparticle scattering rate in Eq. (\ref{QPED-QPSR}) in the SC-state are
respectively reduced as,
\begin{subequations}\label{QPED-QPSR-NS}
\begin{eqnarray}
\varpi_{\bf k}(\omega)&=&\varepsilon_{\bf k}
-{\rm Re}\Sigma_{\rm ph}({\bf k},\omega),\label{QPED-NS}\\
\tilde{\Gamma}_{\bf k}(\omega)&=& {\rm Im}\Sigma_{\rm ph}({\bf k},\omega)
\approx \pi {[2\bar{\Delta}_{\rm PG}({\bf k})]^{2}
\delta(\omega-\varepsilon_{\bf k}}),~~~~~~~~\label{QPSR-NS}
\end{eqnarray}
\end{subequations}
where ${\rm Re}\Sigma_{\rm ph}({\bf k},\omega)$ and
${\rm Im}\Sigma_{\rm ph}({\bf k},\omega)$ are the real and imaginary parts of the
electron normal self-energy $\Sigma_{\rm ph}({\bf k},\omega)$, respectively. The
result in Eq. (\ref{QPSR-NS}) thus indicates directly that there is an intrinsic
link between
the quasiparticle scattering rate and normal-state pseudogap, in agreement with the
ARPES experimental observations\cite{Matt15}. In this case, the location of the EFS
contour in momentum space is determined directly by,
\begin{eqnarray}\label{SC-Fermi-energy}
\varepsilon_{\bf k}+{\rm Re}\Sigma_{\rm ph}({\bf k},0)=0,
\end{eqnarray}
from Eq. (\ref{QPED-NS}), and then the lifetime of the quasiparticles on the EFS
contour is governed by the inverse of the quasiparticle scattering rate
$\tilde{\Gamma}_{\bf k}(0)$ [then the normal-state pseudogap
$\bar{\Delta}_{\rm PG}({\bf k})$] in Eq. (\ref{QPSR-NS}). In other words, the EFS
contour is determined by the poles of the electron propagator at zero energy in
Eq. (\ref{QPED-NS}), while the redistribution of the spectral weight on the EFS
contour is mainly dominated by the quasiparticle scattering rate
$\tilde{\Gamma}_{\bf k}(0)$ [then the normal-state pseudogap
$\bar{\Delta}_{\rm PG}({\bf k})$] in Eq. (\ref{QPSR-NS}). In Fig. \ref{EFS-maps}, we
plot the intensity map of the ARPES spectrum in the normal-state
for the zero binding-energy $\omega=0$ at $\delta=0.10$ with $T=0.002J$. It thus
shows clearly that the spectral weight redistributes along with the EFS contour due
to the opening of the momentum-dependent normal-state pseudogap, where the most
exotic features can be summarized as\cite{Feng16,Feng15a}:
\begin{figure}[h!]
\centering
\includegraphics[scale=0.35]{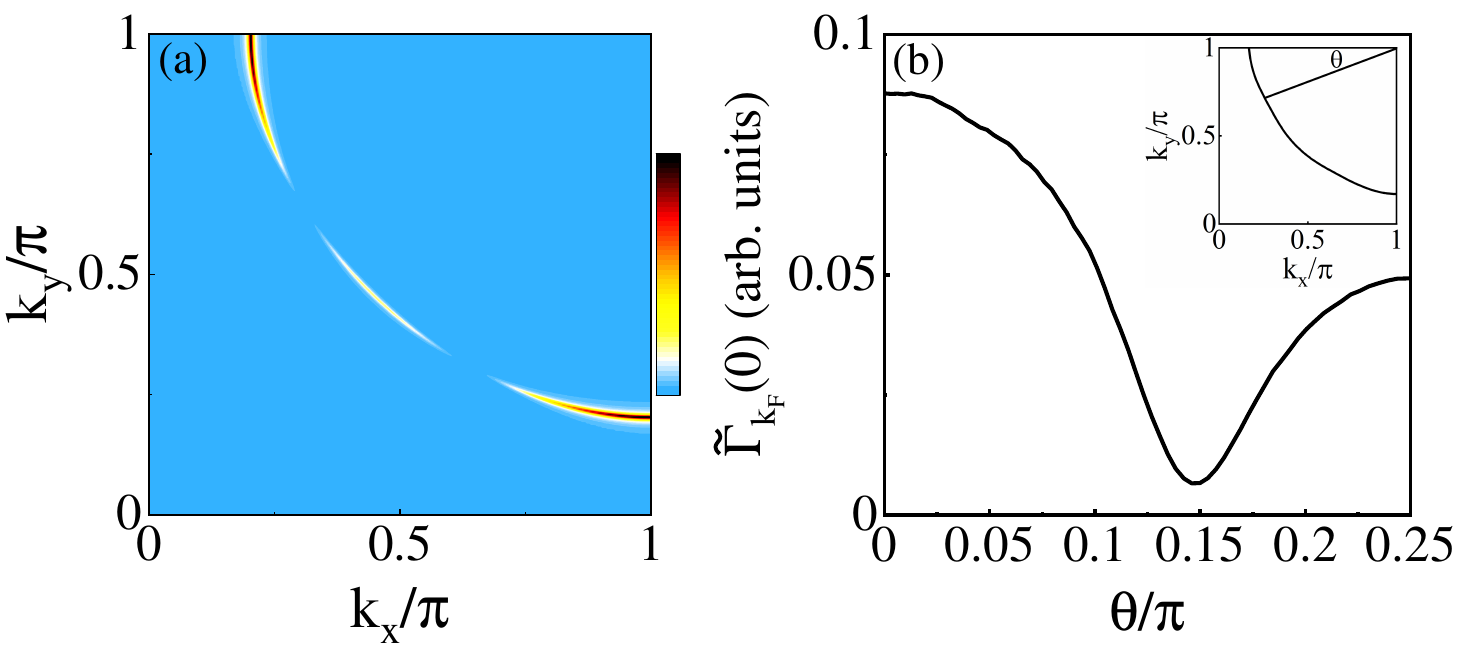}
\caption{(a) The intensity map of the normal-state quasiparticle scattering rate in
the $[k_{x},k_{y}]$ plane and (b) the angular dependence of the normal-state
quasiparticle scattering rate along ${\bf k}_{\rm F}$ from the antinode to the node
for the binding-energy $\omega=0$ at $\delta=0.10$ with $T=0.002J$.
\label{scattering-rate-PG}}
\end{figure}
(a) at around the
antinodal region, the normal-state pseudogap suppresses strongly the spectral weight,
which leads to that the EFS contour at around the antinodal region becomes
unobservable in experiments
\cite{Norman98,Yoshida06,Tanaka06,Kanigel07,Shi08,Sassa11,Horio16,Loret18,Comin14};
(b) at around the nodal region, the normal-state pseudogap suppresses modestly the
spectral weight, which leads to that the EFS contour is broken up into four disconnected
Fermi arcs located at around the nodal region
\cite{Norman98,Yoshida06,Tanaka06,Kanigel07,Shi08,Sassa11,Horio16,Loret18,Comin14};
(c) at around the tips of the Fermi arcs, the renormalization from the quasiparticle
scattering further reduces the most of the spectral weight on the Fermi arcs to the
tips of the Fermi arcs\cite{Shi08,Sassa11,Horio16,Loret18,Comin14}, indicating that the
quasiparticles at around the tips of the disconnected Fermi arcs have the largest
density of states, and then these quasiparticles at around the tips of the Fermi arcs
contribute effectively to the quasiparticle scattering process. It should be noted
that the result of the intensity map of the ARPES spectrum in Fig. \ref{EFS-maps} is
qualitatively consistent with some ARPES
observations\cite{Shi08,Sassa11,Horio16,Loret18,Comin14}, where the sharp quasiparticle
peak with large spectral weight appears always at off-node place. On the other hand,
the ARPES observations\cite{Chatterjee06} also indicate that the highest intensity
points in the intensity map of the ARPES spectrum for {\it a finite binding-energy}
are located exactly at the tips of the Fermi arcs.

Since the redistribution of the spectral weight on the EFS contour is mainly dominated
by the quasiparticle scattering rate $\tilde{\Gamma}_{\bf k}(0)$ [then the normal-state
pseudogap $\bar{\Delta}_{\rm PG}({\bf k})$] in Eq. (\ref{QPSR-NS}) as mentioned above,
the formation of four disconnected Fermi arcs due to the EFS reconstruction is directly
linked with the opening of the highly anisotropic momentum-dependence of the
normal-state pseudogap\cite{Feng16,Feng15a,Matt15}. To see this picture more clearly,
we have made a series of the calculations for the quasiparticle scattering rate
$\tilde{\Gamma}_{{\bf k}_{\rm F}}(0)={\rm Im}\Sigma_{\rm ph}({\bf k}_{\rm F},0)$ in the
normal-state, and the result of the intensity map of
$\tilde{\Gamma}_{{\bf k}_{\rm F}}(0)$ in the $[k_{x},k_{y}]$ plane for the zero
binding-energy
$\omega=0$ at $\delta=0.10$ with $T=0.002J$ is plotted in Fig. \ref{scattering-rate-PG}a.
Apparently, $\tilde{\Gamma}_{{\bf k}_{\rm F}}(0)$ is strong dependence of momentum,
reflecting a fact that the anisotropic $\tilde{\Gamma}_{{\bf k}_{\rm F}}(0)$ [then the
normal-state pseudogap $\bar{\Delta}_{\rm PG}({\bf k}_{\rm F})$] in momentum-space is
strongly non-monotonous angle dependent. To illuminate this fact more clearly, we plot
the angular dependence of $\tilde{\Gamma}_{{\bf k}_{\rm F}}(0)$ along the EFS contour
from the antinode to the node for the zero binding-energy $\omega=0$ at $\delta=0.10$
with $T=0.002J$ in Fig. \ref{scattering-rate-PG}b, where the actual minimum
of $\tilde{\Gamma}_{{\bf k}_{\rm F}}(0)$ always appears at the off-node place, and is
accommodated exactly at the tip of the disconnected Fermi arc. In particular, the
largest magnitude of $\tilde{\Gamma}_{{\bf k}_{\rm F}}(0)$ appears at the antinode,
which is consistent with the experimental
observations\cite{Shi08,Sassa11,Horio16,Loret18,Gh12,Comin14,Neto14,Hash15}, where the
strongest quasiparticle scattering appeared at the antinodes has been also observed.
However, the magnitude of $\tilde{\Gamma}_{{\bf k}_{\rm F}}(0)$ decreases
with the shift of the momentum away from the antinode, leading to that the magnitude
of $\tilde{\Gamma}_{{\bf k}_{\rm F}}(0)$ at around the nodal region is always less
than that at around the antinodal region. This highly anisotropic momentum-dependence
of the quasiparticle scattering rate $\tilde{\Gamma}_{{\bf k}_{\rm F}}(0)$ [then the
normal-state pseudogap $\bar{\Delta}_{\rm PG}({\bf k}_{\rm F})$] thus reduces strongly
the spectral weight on the EFS contour at around the antinodal region, but it has a
more modest effect on the spectral weight on the EFS contour at around the nodal
region, and then the most of the spectral weight on the EFS contour sites at around
the tips of the disconnected Fermi arcs. This special momentum
dependence of $\tilde{\Gamma}_{{\bf k}_{\rm F}}(0)$ [then the normal-state pseudogap
$\bar{\Delta}_{\rm PG}({\bf k}_{\rm F})$] leads to that the density of states of the
quasiparticles at around the nodal region is higher than that at around the antinodal
region, and therefore generates the formation of four disconnected Fermi arcs with the
largest density of states of the quasiparticles located at around the tips of the
disconnected Fermi arcs.

In the past several years, some experimental observations have indicated clearly
that the charge-order state emerges just below the normal-state pseudogap crossover
temperature $T^{*}$, with a characteristic charge-order wave vector that matches well
with the tips on the straight Fermi arcs\cite{Comin14,Gh12,Neto14,Hash15,Comin16}.
Moreover, these experimental observations also show that the charge-order state is
particularly obvious in the underdoped regime, and then the magnitude of the
charge-order wave vector $Q_{\rm CO}$ smoothly decreases with the increase of
doping\cite{Comin14,Gh12,Neto14,Hash15,Comin16}. In the present study, we find that
the theoretical result of the quasiparticle scattering wave vector between the tips
on the straight Fermi arcs shown in Fig. \ref{EFS-maps} at $\delta=0.10$ is
$Q_{\rm CO}=0.295$ (here we use the reciprocal units), which is in qualitatively
agreement with the experimental average value\cite{Comin14,Gh12,Neto14,Hash15,Comin16}
of the charge-order wave vector $Q_{\rm CO}\approx 0.265$ observed in the underdoped
pseudogap phase. The charge-order state is characterized by the charge-order wave
vector, and the qualitative agreement between theory and experiments in the
charge-order wave vector therefore is important to confirm that the charge-order state
is driven by the EFS instability\cite{Feng16}. In particular, the decrease of the
normal-state pseudogap with the increase of doping shown in Fig. \ref{EPG-T-doping}a
also leads to that the charge-order wave vector decreases with the increase of
doping\cite{Feng16}, also in qualitative agreement with the experimental
observations\cite{Comin14,Gh12,Neto14,Hash15,Comin16}. Moreover, this
charge-order correlation developed in the normal-state pseudogap phase can persists
into the SC-state, leading to a coexistence of charge order and superconductivity
below $T_{\rm c}$. The above results therefore indicate that the charge-order state
is manifested within the normal-state pseudogap phase, then the formation of the
disconnected Fermi arcs, the charge-order state, and the normal-state pseudogap in the
underdoped cuprate superconductors are intimately related to each other, and they have
a root in common originated from the interaction between electrons by the exchange of
the spin excitation.

\subsection{Line-shape in energy distribution curve}\label{LSQES}

We now turn to discuss the complicated line-shape in the energy distribution curve
of the underdoped cuprate superconductors. One of the most remarkable
features in the energy distribution curve of the underdoped cuprate superconductors
observed from ARPES experiments has been the PDH structure, where a very sharp peak
appears at the low binding-energy, followed by a dip and then a hump in the higher
binding-energy, leading to the formation of a PDH structure
\cite{Dessau91,Campuzano99,Lu01,Sato02,Borisenko03,Wei08,Sakai13,Hashimoto15,Loret17,DMou17}.
This striking PDH structure was found firstly in
Bi$_{2}$Sr$_{2}$CaCu$_{2}$O$_{8+\delta}$ at around the antinodal region
\cite{Dessau91,Campuzano99}.
Later, this outstanding PDH structure was also observed in other families of cuprate
superconductors\cite{Lu01,Sato02,Borisenko03,Wei08}, and now is a hallmark of the
spectral line-shape of the energy distribution curves in the underdoped cuprate
superconductors. After intensive studies for about four decades, although it is
believed widely that the appearance of the {\it dip} in the energy distribution
curves is due to the very strong scattering of the electrons by {\it a particular
bosonic excitation}, the nature of this bosonic excitation remains controversial,
where two main proposals are disputing the explanations of the origin of the dip
\cite{Dessau91,Campuzano99,Lu01,Sato02,Borisenko03,Wei08,Sakai13,Hashimoto15,Loret17,DMou17}:
in one of the proposals, the dip is associated with the strong electron-phonon
coupling, while in the other, the dip is related to the pseudogap originating from
the spin excitation. In this subsection, we show clearly that the complicated
line-shape in the energy distribution curves is affected strongly by the normal-state
pseudogap, and then the dip in the energy distribution curves associated with the
renormalization of the electrons is electronic in nature\cite{Gao18,Liu21,Cao22}.

\begin{figure}[h!]
\centering
\includegraphics[scale=0.30]{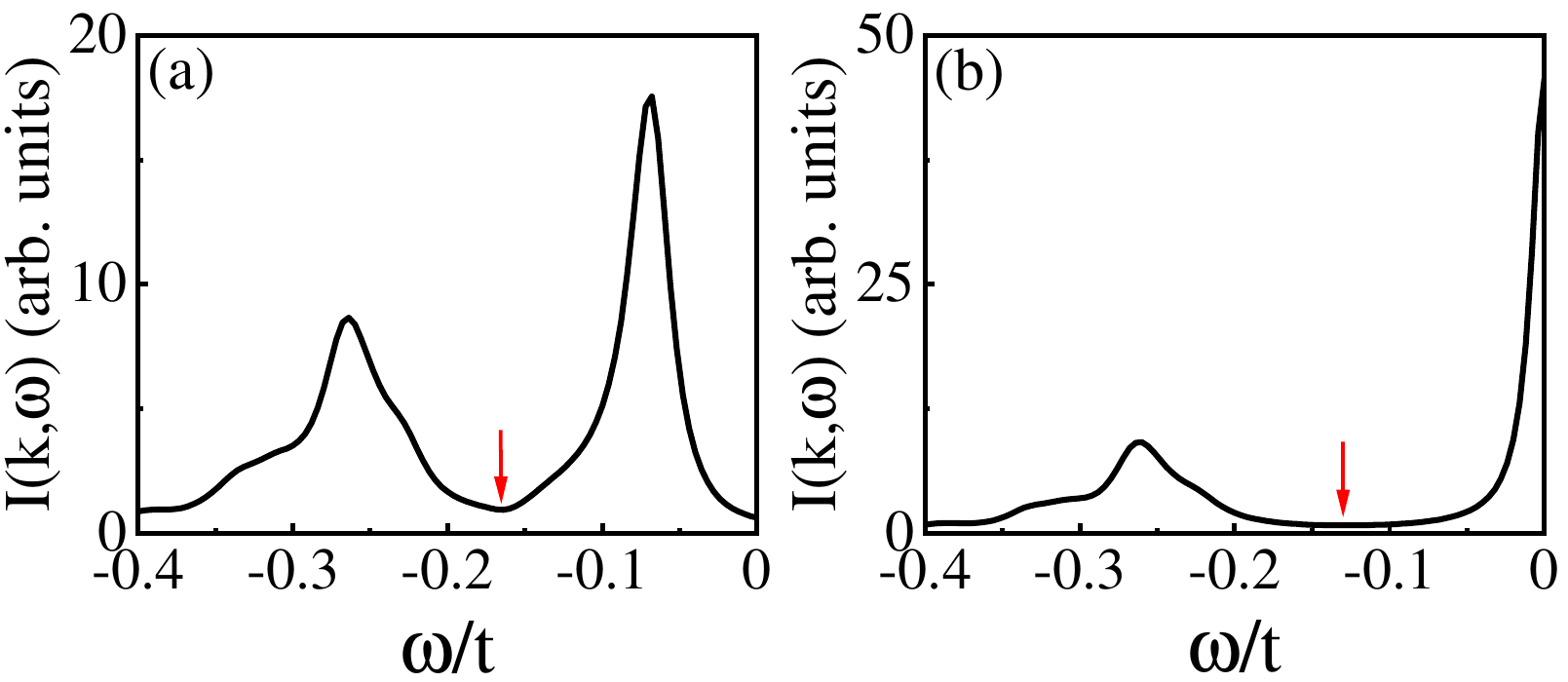}
\caption{The energy distribution curve at the antinode (a) in the
superconducting-state and (b) in the normal-state as a function of binding-energy
for $\delta=0.10$ with $T=0.002J$. The red arrow indicating the position of the dip
at the dip energy $\omega_{\rm dip}$. \label{PDH-PG}}
\end{figure}

In Fig. \ref{PDH-PG}a, we plot the energy distribution curve at the antinode
in the SC-state as a function of binding-energy for $\delta=0.10$ with $T=0.002J$,
where the energy distribution curve consists of two distinct peaks: a very sharp
low-energy peak and a relatively broad high-energy peak. The high-energy peak is
corresponding to the hump, and the low-energy peak is corresponding to the SC
quasiparticle excitation, while the deviation of the low-energy peak from EFS is due
to the opening of the d-wave type SC gap. A spectral dip is located between these two
distinct peaks, which is corresponding to the intensity depletion region. The total
contributions for the energy distribution curve thus lead to the formation of the
PDH structure, in qualitative agreement with the ARPES experimental observations on
cuprate superconductors
\cite{Dessau91,Campuzano99,Lu01,Sato02,Borisenko03,Wei08,Sakai13,Hashimoto15,Loret17,DMou17}.
However, this exotic PDH structure is also observed experimentally in the underdoped
pseudogap phase\cite{Campuzano99,Hashimoto15}.
To see this point more clearly, the energy distribution curve
at the antinode in the normal-state as a function
of binding-energy for $\delta=0.10$ with $T=0.002J$ is also plotted in
Fig. \ref{PDH-PG}b. Comparing it with Fig. \ref{PDH-PG}a for the same set of
parameters except for $\bar{\Delta}=0$, we see that (i) the overall feature of the
line-shape in the energy distribution curve in the normal-state presents a
similar behavior of the line-shape in the SC-state; (ii) however, in a difference
to the case in the SC-state, the sharp low-energy peak is shifted to EFS in the
normal-state due to the disappearance of the SC gap, also in qualitative agreement
with the ARPES experimental observations\cite{Campuzano99,Hashimoto15}.

\begin{figure}[h!]
\centering
\includegraphics[scale=0.053]{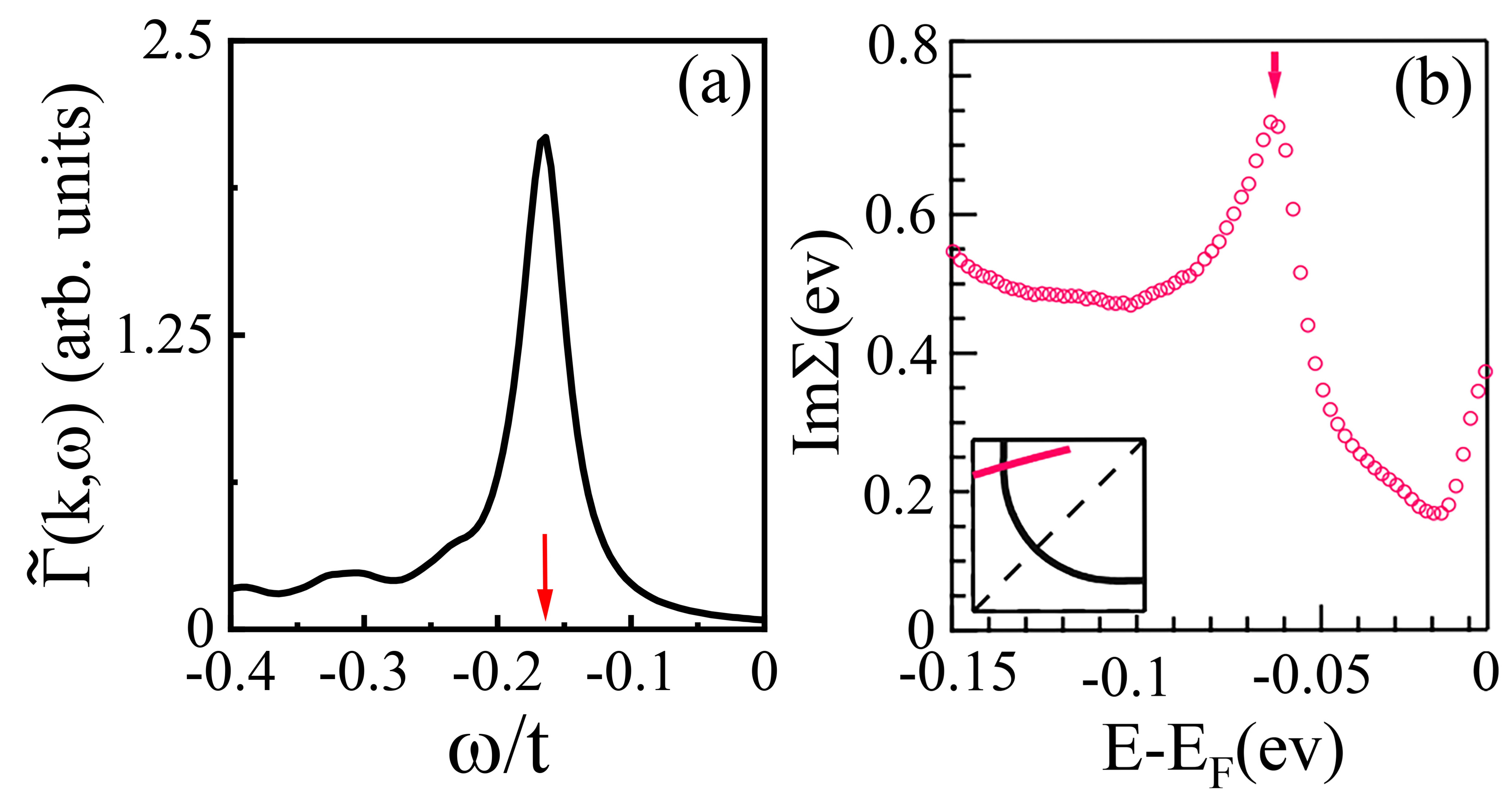}
\caption{(a) The quasiparticle scattering rate at the antinode in the
superconducting-state
as a function of binding-energy for $\delta=0.10$ with $T=0.002J$. (b) The
experimental result of Bi$_{2}$Sr$_{2}$CaCu$_{2}$O$_{8+\delta}$ taken from
Ref. \onlinecite{DMou17}.\label{scattering-rate-energy}}
\end{figure}

In the energy distribution curve\cite{Damascelli03,Campuzano04,Fink07}, a
quasiparticle with a long lifetime is observed as a sharp peak in intensity, and a
quasiparticle with a short lifetime is observed as a broad hump, while the dip in
the energy distribution curve implies that the quasiparticle scattering rate has a
sharp peak accommodated at around the dip energy\cite{DMou17,Gao18,Liu21,Cao22}.
This can be well understood from the quasiparticle energy dispersion
$\varpi_{\bf k}(\omega)$ in Eq. (\ref{QPED}) and quasiparticle scattering rate
$\tilde{\Gamma}_{\bf k}(\omega)$ in Eq. (\ref{QPSR}). The ARPES spectrum
exhibits a peak when the incoming photon energy $\omega$ is equal to the
quasiparticle excitation energy $\varpi_{\bf k}(\omega)$, i.e.,
\begin{eqnarray}\label{band}
\omega-\varpi_{\bf k}(\omega)=\omega-\varepsilon_{\bf k}
-{\rm Re}\Sigma_{\rm tot}({\bf k},\omega)=0,
\end{eqnarray}
and then the width of this peak at the energy $\omega$ is dominated by the
quasiparticle scattering rate $\tilde{\Gamma}_{\bf k}(\omega)$ [then the normal-state
pseudogap $\bar{\Delta}_{\rm PG}({\bf k})$].
In Fig. \ref{scattering-rate-energy}a,
we plot $\tilde{\Gamma}_{\bf k}(\omega)$ at the antinode in the SC-state as a
function of binding-energy for $\delta=0.10$ with $T=0.002J$. For a clear comparison,
the experimental result\cite{DMou17} detected in
Bi$_{2}$Sr$_{2}$CaCu$_{2}$O$_{8+\delta}$ in the SC-state is also shown in
Fig. \ref{scattering-rate-energy}b.
\begin{figure}[h!]
\centering
\includegraphics[scale=0.20]{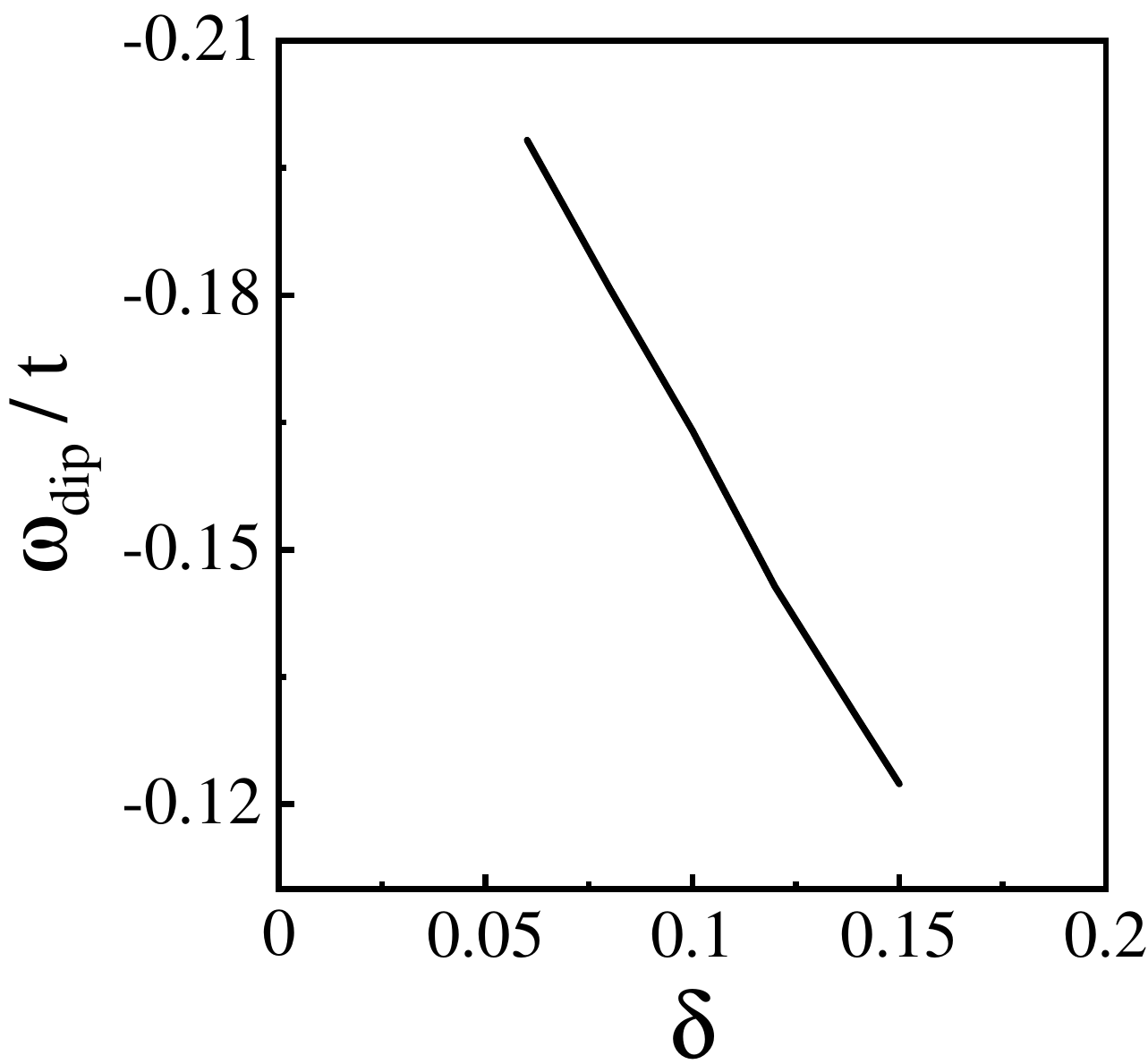}
\caption{The antinodal dip energy $\omega_{\rm dip}$ in the superconducting-state as
function of doping with $T=0.002J$. \label{dip-energy-doping}}
\end{figure}
The present result in
Fig. \ref{scattering-rate-energy}a shows that $\tilde{\Gamma}_{\bf k}(\omega)$ has a
well-pronounced peak structure\cite{Gao18,Liu21,Cao22}, where
$\tilde{\Gamma}_{\bf k}(\omega)$ exhibits a sharp peak at the binding-energy
$\omega_{\rm dip}$, and then the weight of this sharp peak reduces rapidly away from
the peak, which is well consistent with the experimental results\cite{DMou17}.
In particular, the position of this sharp peak
in $\tilde{\Gamma}_{\bf k}(\omega)$ is just corresponding to the position of the
dip in the PDH structure shown in Fig. \ref{PDH-PG}b, which therefore confirms that
the intensity depletion in the energy distribution curve at around the dip is a
natural consequence of the peak structure in $\tilde{\Gamma}_{\bf k}(\omega)$.
Moreover, the dip in the energy distribution curve is shifted towards to EFS when
the doping is increased in the underdoped regime. To see this point more clearly,
we plot the dip energy $\omega_{\rm dip}$ in the SC-state as a function of doping
with $T=0.002J$ in Fig. \ref{dip-energy-doping}, where $\omega_{\rm dip}$ as a
function of doping in the underdoped regime presents a similar behavior of the
normal-state pseudogap $\bar{\Delta}_{\rm PG}$ shown in Fig. \ref{EPG-T-doping}a,
also in qualitative agreement with the experimental observations
\cite{Dessau91,Campuzano99,Lu01,Sato02,Borisenko03,Wei08,Sakai13,Hashimoto15,Loret17,DMou17}.
This SC-state coexists with the normal-state pseudogap below $T_{\rm c}$. In this
case, as a comparison, we have also calculated the quasiparticle scattering rate
$\tilde{\Gamma}_{\bf k}(\omega)$ at the antinode in the normal-state pseudogap phase
as a function of binding-energy for the same set of parameters as in
Fig. \ref{scattering-rate-energy}a, and the obtained result shows that the overall
feature of the quasiparticle scattering rate in the normal-state pseudogap
phase is almost the same as that in the SC-state shown in
Fig. \ref{scattering-rate-energy}a, and then the peak structure in the quasiparticle
scattering rate leads to the formation of a similar PDH structure in the energy
distribution curve in the normal-state pseudogap phase as shown in
Fig. \ref{PDH-PG}a. The above results therefore also show that the well-known PDH
structure in the energy distribution curve can be attributed to the emergence of the
normal-state pseudogap.

\subsection{ARPES autocorrelation}\label{ARPES-autocorrelation}

The ARPES autocorrelation\cite{Chatterjee06,He14,Restrepo23},
\begin{equation}\label{ACF}
{\bar C}({\bf q},\omega)={1\over N}\sum_{\bf k}I({\bf k}+{\bf q},\omega)
I({\bf k},\omega), ~~~~~~~
\end{equation}
can be analyzed in terms of the ARPES spectrum (\ref{QPES}), where
the summation of momentum ${\bf k}$ is restricted within the first Brillouin zone (BZ).
This ARPES autocorrelation therefore detects the autocorrelation of the ARPES spectra
at two different momenta ${\bf k}$ and ${\bf k}+{\bf q}$ with a
fixed energy $\omega$. Experimentally, the ARPES spectrum reveals straightforwardly
the momentum-space electronic structure\cite{Damascelli03,Campuzano04,Fink07}, while
the ARPES autocorrelation spectrum reveals straightforwardly the effectively
momentum-resolved joint density of states in the electronic
state\cite{Chatterjee06,He14,Restrepo23}, and then both the spectra yield the important
insights into the nature of the electronic state. More importantly, the experimental
observations have provided the unambiguous evidence\cite{Chatterjee06,He14,Restrepo23}
that the sharp peaks in the ARPES autocorrelation spectrum in the SC-state are directly
linked to the quasiparticle scattering wave vectors ${\bf q}_{i}$ connecting the tips
of the Fermi arcs obeying an octet scattering model, and are also well consistent with
the quasiparticle scattering interference peaks observed from the Fourier transform STS
experiments
\cite{Hoffman02,McElroy03,Yin21,Pan00,Kohsaka08,Hanaguri09,Lee09,Vishik09,Schmidt11}.
However, although the experimental data of the ARPES autocorrelation spectrum in the
SC-state have been available\cite{Chatterjee06,He14,Restrepo23}, the experimental
data of the ARPES autocorrelation spectrum in the normal-state pseudogap phase are
still lacking to date. In this subsection, we further discuss the influence of the
pseudogap on the electronic state of the underdoped cuprate superconductors in terms
of the ARPES autocorrelation.

\begin{figure}[h!]
\centering
\includegraphics[scale=0.30]{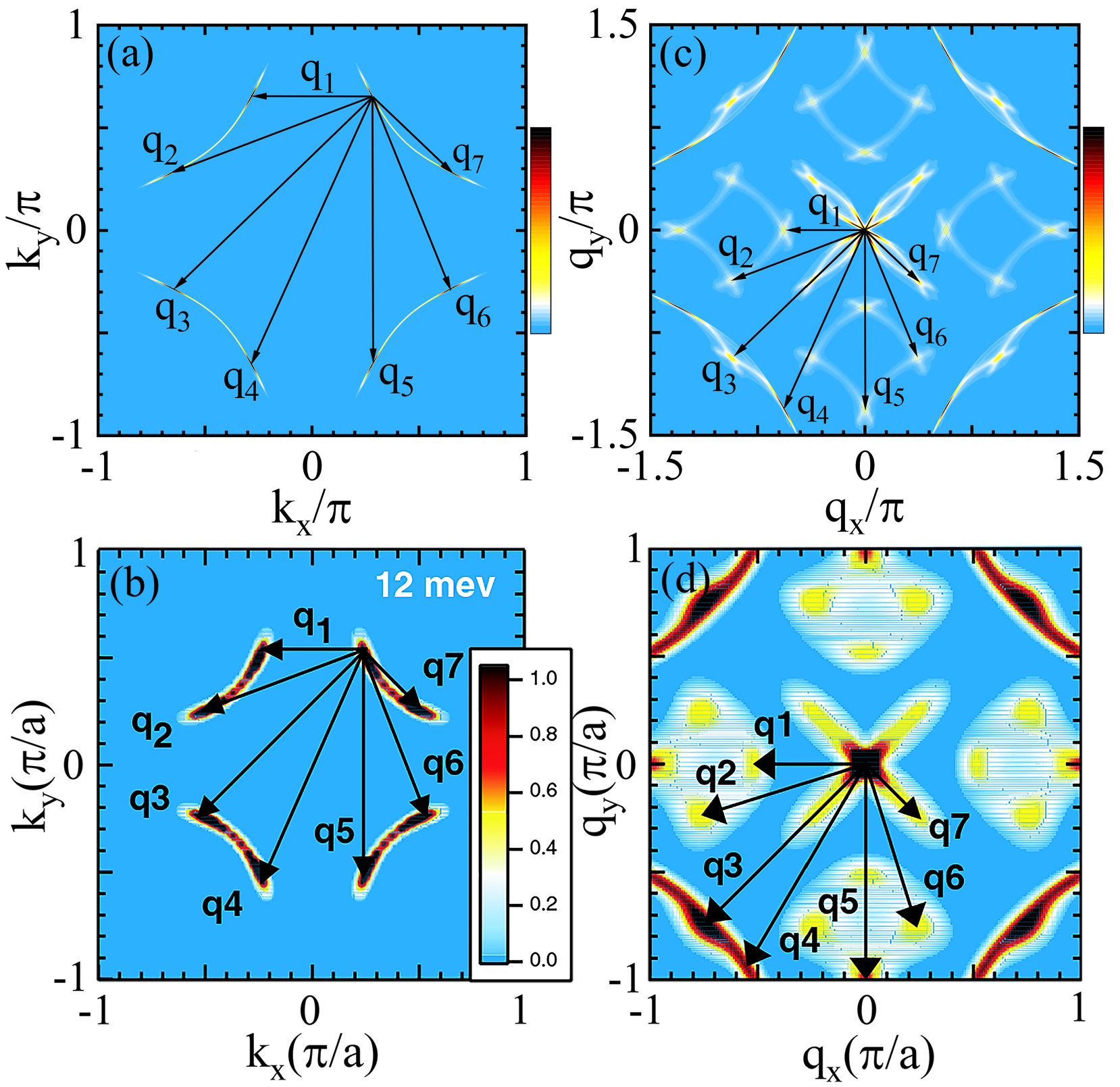}
\caption{(Color online) (a) The intensity map of the ARPES spectrum in the
superconducting-state for the binding-energy $\omega=0.12J$ at $\delta=0.10$
with $T=0.002J$. (b) The experimental result observed on
Bi$_{2}$Sr$_{2}$CaCu$_{2}$O$_{8+\delta}$ for the binding-energy $\omega=12$ meV
taken from Ref. \onlinecite{Chatterjee06}. (c) The intensity map of the ARPES
autocorrelation in the superconducting-state for the binding-energy $\omega=0.12J$
at $\delta=0.10$ with $T=0.002J$. (d) The experimental result observed from
Bi$_{2}$Sr$_{2}$CaCu$_{2}$O$_{8+\delta}$ for the binding-energy $\omega=12$ meV
taken from Ref. \onlinecite{Chatterjee06}. The eight tips of the disconnected Fermi
arcs determine the scattering within the octet scattering model, while the scattering
wave vectors which connect these tips of the disconnected Fermi arcs are shown as
arrows labeled by the designation of each scattering wave vector ${\bf q}_{i}$.
\label{CEC-map}}
\end{figure}

In subsection \ref{EFSR-NS}, the intensity map of the normal-state ARPES spectrum at
the zero binding-energy has been discussed\cite{Feng16}, where the redistribution of
the spectral weight on EFS leads to the formation of the disconnected Fermi arcs.
However, this spectral weight redistribution shown in Fig. \ref{EFS-maps} at the case
for the zero binding-energy can persist into the system at the case for a finite
binding-energy\cite{Gao19}. For a convenience in the following discussions of the
unusual features of the ARPES autocorrelation, we firstly discuss the essential
properties of the constant energy contour at the case for a finite binding-energy. In
Fig. \ref{CEC-map}a, we plot an intensity map of the SC-state ARPES spectrum
$I({\bf k},\omega)$ in Eq. (\ref{QPES}) for the binding-energy $\omega=0.12J$ at
$\delta=0.10$ with $T=0.002J$. For a clear comparison, the ARPES experimental
data\cite{Chatterjee06} detected from Bi$_{2}$Sr$_{2}$CaCu$_{2}$O$_{8+\delta}$ in the
SC-state for the binding-energy $\omega=12$ meV is also shown in Fig. \ref{CEC-map}b.
As in the case of the spectral weight redistribution shown in Fig. \ref{EFS-maps},
(i) the formation of the disconnected Fermi arcs on EFS induced by the normal-state
pseudogap persists into the case of the constant energy contour for a finite
binding-energy\cite{Gao19}, where the notation {\it Fermi arcs} on the constant energy
contour has been used even for a finite binding-energy\cite{Chatterjee06}; (ii)
the points with the highest intensity do not accommodated at around the nodes, but
they locate exactly at around the tips of the disconnected Fermi arcs; (iii) these
tips of the disconnected Fermi arcs connected by the scattering wave vectors
${\bf q}_{i}$ construct an {\it octet scattering model}, which is well consistent with
the corresponding ARPES experimental observations\cite{Chatterjee06,He14,Restrepo23}.

We are now ready to discuss the ARPES autocorrelation in the underdoped cuprate
superconductors. In Fig. \ref{CEC-map}c, we plot the intensity map of the ARPES
autocorrelation ${\bar C}({\bf q},\omega)$ in the SC-state for the binding-energy
$\omega=0.12J$ at $\delta=0.10$ with $T=0.002J$ in comparison with (d) the
experimental result\cite{Chatterjee06} of the intensity map of the ARPES
autocorrelation detected from Bi$_{2}$Sr$_{2}$CaCu$_{2}$O$_{8+\delta}$ in the
SC-state for the
binding-energy $\omega=12$ meV. The result in Fig. \ref{CEC-map}c thus shows that
some sharp peaks appear in the ARPES autocorrelation pattern, where the joint
density of states is highest\cite{Gao19}.
\begin{figure}[h!]
\centering
\includegraphics[scale=0.33]{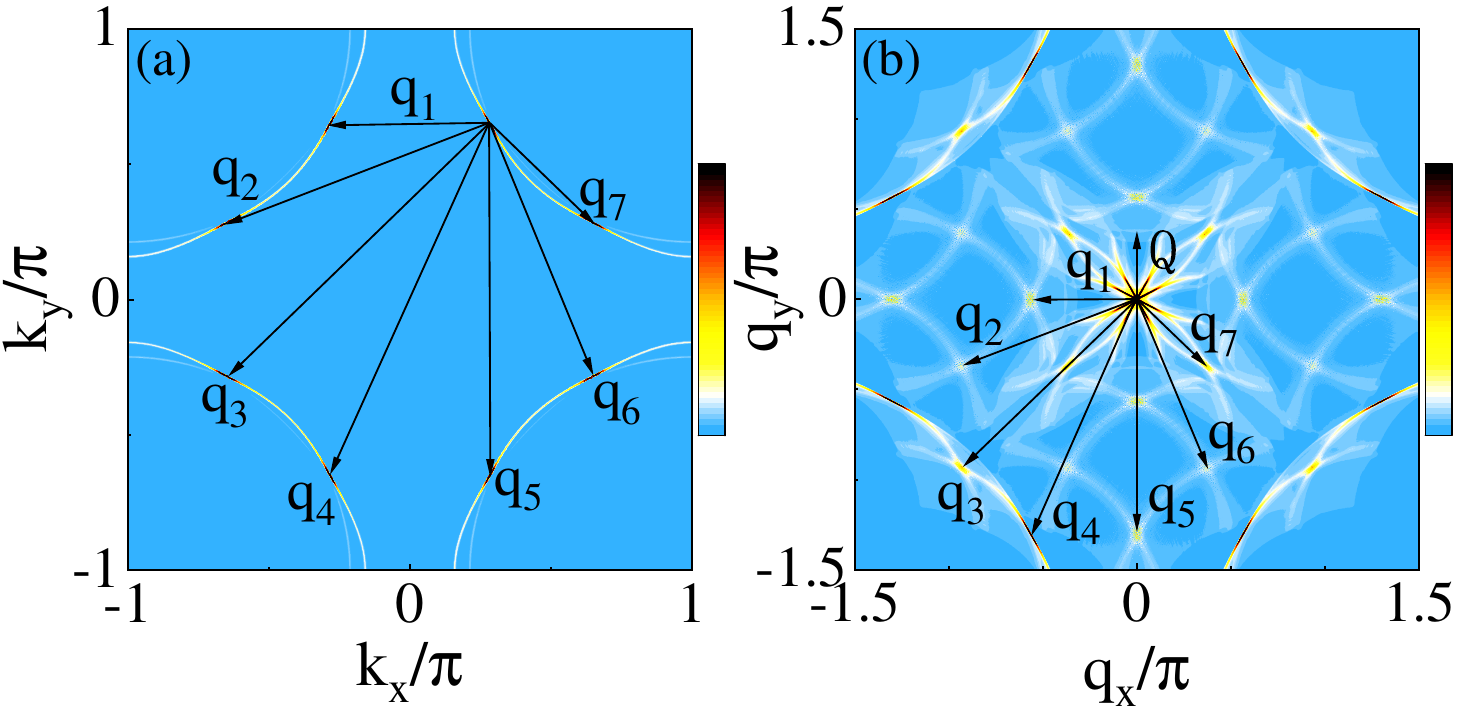}
\caption{(Color online) (a) The intensity map of the ARPES spectrum in the
normal-state for
the binding-energy $\omega=12$ meV at $\delta=0.10$ with $T=0.002J$. (b) The
intensity map of the ARPES autocorrelation in the normal-state for the
binding-energy $\omega=0.12J$ at $\delta=0.10$ with $T=0.002J$. As the case of the
superconducting-state shown in Fig. \ref{CEC-map}, the scattering wave vectors
${\bf q}_{i}$ in the normal-state also obey the octet scattering model, while
${\bf Q}$ in (b) indicates the checkerboard scattering wave vector.
\label{CEC-map-NS}}
\end{figure}
More specifically, these sharp peaks in
the ARPES autocorrelation pattern are straightforwardly correlated with the wave
vectors ${\bf q}_{i}$ connecting the tips of the disconnected Fermi arcs on the
constant energy contour shown in Fig. \ref{CEC-map}a, which are well consistent with
the ARPES experimental observations\cite{Chatterjee06,He14,Restrepo23}. The above
results in Fig. \ref{CEC-map} thus also confirm that in the SC-state, the
{\it octet scattering model} constructed by the eight tips of the disconnected Fermi
arcs can give a consistent description of the regions of the highest joint density
of states in the ARPES autocorrelation spectrum\cite{Gao19}.

Although the SC-state of the underdoped cuprate superconductors coexists and
competes with the normal-state pseudogap state, the quasiparticle still exhibits
the particle-hole mixing similar to that in the conventional superconductors
\cite{Damascelli03,Campuzano04,Fink07}. In particular, in the low-energy limit, the
quasiparticle energy dispersion $\varpi_{\bf k}(\omega)$ in Eq. (\ref{QPED}) in the
SC-state can be reduced as [see Appendix \ref{Derivation-of-propagators}],
$\varpi_{\bf k}=E_{\bf k}$, where $E_{\bf k}=\sqrt{\bar{\varepsilon}^{2}_{\bf k}
+\mid\bar{\Delta}_{{\rm Z}{\bf k}}\mid^{2}}$, with $\bar{\varepsilon}_{\bf k}
=Z_{\rm F}\varepsilon_{\bf k}$, and
$\bar{\Delta}_{{\rm Z}{\bf k}}=Z_{\rm F}\bar{\Delta}_{\bf k}$,
while the single-particle coherent weight $Z_{\rm F}$ that is directly associated
with the normal-state pseudogap has been given explicitly in
Appendix \ref{Derivation-of-propagators}. In this case, the scattering wave vector
${\bf q}_{i}$ dominates the ARPES autocorrelation at energy $\omega$, since these
${\bf k}$-pairs on the constant energy contour shown in Fig. \ref{CEC-map}a
connected by the wave vector ${\bf q}_{i}$ have a high joint density of states.
This also follows from a basic fact that the magnitude of the SC gap
$\mid\bar{\Delta}_{\bf k}\mid$ is ${\bf k}$-dependent, and then some different minima
$\omega_{i}$, at which the quasiparticles emerge, occur at the tips of the
disconnected Fermi arcs\cite{Hoffman02,McElroy03}. On the other hand, the
square-lattice CuO$_{2}$ plane has a four-fold symmetry, this implies that at any
non-zero energy $\omega_{i}$, there are only eight possible ${\bf k}_{\rm F}$ values
at which $\mid\bar{\Delta}_{{\rm Z}{\bf k}}\mid=\omega_{i}$ in the first BZ
\cite{Hoffman02,McElroy03}. The highest joint density of states for the quasiparticle
scattering at this energy $\omega_{i}$ occur at ${\bf q}_{i}$-vectors connecting the
eight tips of the Fermi arcs. This is why the sharp peaks in the ARPES autocorrelation
occur at these ${\bf q}_{i}$-vectors. However, in the normal-state pseudogap phase,
where the SC gap $\bar{\Delta}=0$, the additional distinct peaks (then the additional
distinct electronic orders) in the ARPES autocorrelation spectrum that are suppressed
strongly by the SC gap appear. To understand the nature of these additional distinct
peaks in the ARPES autocorrelation spectrum in the normal-state more clearly, we plot
(a) the intensity map of the ARPES spectrum $I({\bf k},\omega)$ in the normal-state
and (b) the intensity map of the ARPES autocorrelation ${\bar C}({\bf q},\omega)$ in
the normal-state for the binding-energy $\omega=12$ meV at $\delta=0.10$ with
$T=0.002J$ in Fig. \ref{CEC-map-NS}. In a careful comparison with the corresponding
results in the SC-state shown in Fig. \ref{CEC-map}, the similarities and differences
of the ARPES autocorrelation patterns between the SC- and normal-states can be
summarized as: (i) the overall features of the redistribution of the spectral weight
on the constant-energy-contour for a finite binding-energy in the SC-state persist
into the case in the normal-state for a finite binding-energy, indicating that the
redistribution of the spectral weight on the constant-energy-contour is mainly
dominated by the normal-state pseudogap; (ii) the overall
features of the ARPES autocorrelation peaks labelled by ${\bf q}_{i}$ ($i=1,...,7$)
in the normal-state are well consistent with those in the SC-state, reflecting an
experimental fact that the multiple electronic order-states that emerge in the
normal-state above $T_{\rm c}$ persist into the SC-state below $T_{\rm c}$, and then
they coexist and compete with the SC-state. In particular, the quasiparticle scattering
between two tips of the straight Fermi arcs with the scattering wave vector
${\bf q}_{1}={\bf Q}_{\rm CO}$ matches well with the corresponding charge-order wave
vector found in the resonant X-ray scattering measurements and STS experiments
\cite{Comin14,Gh12,Neto14,Hash15,Comin16,Campi15}, which thus confirms the
charge-order state is driven by the EFS instability
\cite{Comin14,Gh12,Neto14,Hash15,Comin16,Feng16}; (iii) however, the additional
distinct peaks along the parallel direction that are suppressed strongly by the
SC-gap emerge, leading to the formation of the {\it checkerboard} peaks (then the
{\it checkerboard charge ordering}), with the scattering wave vectors of
${\bf Q}\approx [\pm 0.38\pi,0]$ and ${\bf Q}\approx [0,\pm 0.38\pi]$. In
particular, it should be noted that although the experimental data of the
checkerboard charge ordering in the underdoped pseudogap phase observed directly
from the ARPES autocorrelation are still lacking to date, the early ARPES
experiments\cite{Shen05} have observed the {\it checkerboard charge ordering} in
the underdoped pseudogap phase above $T_{\rm c}$. Moreover, by virtue of systematic
studies using the STS measurements\cite{Vershinin04}, the electronic checkerboard
patterns have been detected directly in the underdoped pseudogap phase above
$T_{\rm c}$. On the other hand, some STS experimental
measurements\cite{Hanaguri04,McElroy05,Wise08} also indicate that although the
electronic checkerboard ordering appeared in the underdoped pseudogap phase can
persist into the SC-state below $T_{\rm c}$, the magnitude of the incoming photon
energy in these experiments is much larger than that of the SC gap
($|\omega|>\bar{\Delta}$). However, whether the checkerboard charge ordering
observed in the normal-state pseudogap phase or observed in the SC-state with
$|\omega|>\bar{\Delta}$, the average
values of the checkerboard charge-order wave vectors have been identified in these
experiments\cite{Hoffman02,McElroy03,Shen05,Vershinin04,Hanaguri04,McElroy05,Wise08}
as ${\bf Q}\approx [\pm 0.44\pi,0]\pm 15\%$ and
${\bf Q}\approx [0,\pm 0.44\pi]\pm 15\%$,
which are qualitatively consistent with the present results.

Since all the features observed in the STS experiments, both in the SC-state and in
the pseudogap phase, are related to the momentum-space profiles of the joint density
of states\cite{McElroy06}, we\cite{Li25} have started very recently from the
homogeneous electron propagator in Eq. (\ref{EDODGF}) to study the local density of
states (LDOS) modulation in the underdoped cuprate superconductors within the
$T$-matrix approach, where LDOS is derived for various kinds of impurities, and the
obtained results show that the pronounced peaks in the momentum-space LDOS modulation
pattern are qualitatively consistent with the present results of the ARPES
autocorrelation patterns shown in Fig. \ref{CEC-map} and Fig. \ref{CEC-map-NS}. These
and the related results will be presented elsewhere.

\section{Summary}\label{Summary}

Within the framework of the kinetic-energy-driven superconductivity, we have studied
the effect of the normal-state pseudogap on the low-energy electronic structure in
the underdoped cuprate superconductors. Our results show that the
interaction between electrons directly from the kinetic energy by the exchange of the
spin excitation generates the normal-state pseudogap-state in the particle-hole
channel and the SC-state in the particle-particle channel, where the normal-state
pseudogap and the SC gap respectively originate from the electron normal self-energy
in the particle-hole channel and the electron anomalous self-energy in the
particle-particle channel, and are derived explicitly by taking into account the
vertex correction. This normal-state pseudogap-state coexists and competes with the
SC-state below $T_{\rm c}$, which leads to that $T_{\rm c}$ is raised gradually with
the increase of doping in the underdoped regime, and then achieves its maximum at
around the optimal doping, subsequently, $T_{c}$ decreases monotonically with the
increase of doping in the overdoped regime. However, in a striking contrast to
$T_{\rm c}$ in the underdoped regime, $T^{*}$ is much higher than $T_{\rm c}$ in the
underdoped regime, then it smoothly decreases when the doping is increased, and
both $T^{*}$ and $T_{c}$ converge to the end of the SC dome. Concomitantly, a number
of the exotic features of the low-energy electronic structure are directly correlated
to the opening of the normal-state pseudogap: (i) the redistribution of the spectral
weight on the EFS contour, where the pseudogap suppresses strongly the spectral
weight of the ARPES spectrum on the EFS contour at around the antinodal region, while
it suppresses modestly the spectral weight at around the nodal region, leading to the
formation of the disconnected Fermi arcs centered around the nodal region; (ii) the
PDH structure in the energy distribution curve, where the emergence of the dip in
the PDH structure is due to the presence of the well-pronounced peak structure in
the quasiparticle scattering rate (then the normal-state pseudogap); (iii) the notable
checkerboard charge ordering in the underdoped pseudogap phase, where the checkerboard
peaks found in the ARPES autocorrelation pattern are intimately related to the opening
of the normal-state pseudogap. Our results therefore indicate that the {\it same spin
excitation} that governs both the normal-state pseudogap-state and the SC-state
naturally generates the exotic features of the low-energy electronic structure in the
underdoped cuprate superconductors.

\section*{Acknowledgements}

X.L. and S.P. are supported by the National Key Research and Development Program of
China under Grant Nos. 2023YFA1406500 and 2021YFA1401803, and the National Natural
Science Foundation of China (NSFC) under Grant No. 12274036. M.Z. is supported by
NSFC under Grant No. 12247116 and the Special Funding for Postdoctoral Research
Projects in Chongqing under Grant No. 2024CQBSHTB3156. H.G. acknowledge support from
NSFC grant Nos. 11774019 and 12074022. This paper is written in honor of Professor
Jan Zaanen, who has made many original works in the field of the high-temperature
superconductivity research.

\begin{appendix}


\section{Derivation of Full electron diagonal and off-diagonal propagators}
\label{Derivation-of-propagators}

In this Appendix, our main goal is to derive the electron normal and anomalous
self-energies in Eq. (\ref{E-N-AN-SE}) of the main text [then the full electron
diagonal and off-diagonal propagators $G({\bf k},\omega)$ and
$\Im^{\dagger}({\bf k},\omega)$ in Eq. (\ref{EDODGF}) of the main text] within the
framework of the kinetic-energy-driven superconductivity by taking into account
the vertex correction. The $t$-$J$ model (\ref{tjham}) in the fermion-spin
representation (\ref{CSS}) can be expressed as\cite{Feng9404,Feng15},
\begin{eqnarray}\label{CSS-tJ-model}
H&=&\sum_{ll'}t_{ll'}(h^{\dagger}_{l'\uparrow}h_{l\uparrow}S^{+}_{l}S^{-}_{l'}
+h^{\dagger}_{l'\downarrow}h_{l\downarrow}S^{-}_{l}S^{+}_{l'})\nonumber\\
&-&\mu_{\rm h}\sum_{l\sigma}h^{\dagger}_{l\sigma}h_{l\sigma}
+J_{\rm eff}\sum_{\langle ll'\rangle}{\bf S}_{l}\cdot {\bf S}_{l'},
\end{eqnarray}
with the charge-carrier chemical potential $\mu_{\rm h}$,
$J_{\rm eff}=(1-\delta)^{2}J$, and the charge-carrier doping concentration
$\delta=\langle h^{\dagger}_{l\sigma}h_{l\sigma}\rangle$. As a natural consequence,
the kinetic-energy term in the fermion-spin theory description of the $t$-$J$ model
(\ref{CSS-tJ-model}) has been converted into the strong coupling between charge and
spin degrees of freedom of the constrained electron, which governs the essential
physics of cuprate superconductors.

\subsection{Mean-field formalism}\label{Mean-field-theory}

In the mean-field (MF) approximation, the $t$-$J$ model (\ref{CSS-tJ-model}) can be
decoupled as\cite{Feng9404,Feng15},
\begin{subequations}\label{MF-t-J-model}
\begin{eqnarray}
H_{\rm MF}&=&H_{\rm t}+H_{\rm J}+H_{0},
\end{eqnarray}
\begin{eqnarray}
H_{\rm t}&=&\sum_{ll'\sigma}t_{ll'}\chi_{ll'}h^{\dagger}_{l'\sigma}h_{l\sigma}
-\mu_{\rm h}\sum_{l\sigma}h^{\dagger}_{l\sigma}h_{l\sigma}, ~~~~~~~~
\label{MF-t-term}\\
H_{\rm J}&=& {1\over 2}J_{\rm eff}\sum_{\langle ll'\rangle}[S^{+}_{l}S^{-}_{l'}
+S^{-}_{l}S^{+}_{l'}+2S^{\rm z}_{l}S^{\rm z}_{l'}]\nonumber\\
&+& \sum_{ll'}t_{ll'}\phi_{ll'}(S^{+}_{l}S^{-}_{l'}+S^{-}_{l}S^{+}_{l'}),~~~~~~
\label{MF-J-term}
\end{eqnarray}
\end{subequations}
where $H_{0}=-2\sum_{ll'}t_{ll'}\chi_{ll'}\phi_{ll'}$, the charge-carrier's
particle-hole parameter
$\phi_{ll'}=\langle h^{\dagger}_{l\sigma}h_{l'\sigma}\rangle$, and the spin
correlation function $\chi_{ll'}=\langle S_{l}^{+}S_{l'}^{-}\rangle$.

With the help of the above charge-carrier part (\ref{MF-t-term}), the MF
charge-carrier propagator is obtained straightforwardly as
\cite{Feng9404,Feng15},
\begin{eqnarray}\label{MFHGF}
g^{(0)}({\bf k},\omega)={1\over \omega-\xi_{\bf k}},
\end{eqnarray}
where the MF charge-carrier energy dispersion is given explicitly as,
\begin{eqnarray}\label{MFCCS}
\xi_{\bf k} = 4\chi_{1}t\gamma_{\bf k}-4\chi_{2}t'\gamma_{\bf k}'-\mu_{\rm h},
\end{eqnarray}
with $\chi_{1}=\langle S_{l}^{+}S_{l+\hat{\eta}}^{-}\rangle$ and
$\chi_{2}=\langle S_{l}^{+}S_{l+\hat{\tau}}^{-}\rangle$.

Now we turn to derive the spin propagator in the MF approximation. In the doped
regime without an antiferromagnetic long-range order (AFLRO), i.e.,
$\langle S^{\rm z}_{l}\rangle =0$, the spin propagator can be calculated in terms of
the Kondo-Yamaji decoupling approximation\cite{Kondo72}, which is a stage one-step
further than the Tyablikov's decoupling approximation\cite{Tyablikov67}. However, in
the MF level, the spin part (\ref{MF-J-term}) is an anisotropic Heisenberg model, we
therefore need two spin propagators $D(l-l',t-t')=-i\theta(t-t')\langle
[S^{+}_{l}(t),S^{-}_{l'}(t')]\rangle=\langle\langle S^{+}_{l}(t);S^{-}_{l'}(t')
\rangle\rangle$ and $D_{\rm z}(l-l',t-t')=-i\theta(t-t')\langle
[S^{\rm z}_{l}(t),S^{\rm z}_{l'}(t')]\rangle=\langle\langle
S^{\rm z}_{l}(t);S^{\rm z}_{l'}(t')\rangle\rangle$ to give a proper description of
the essential properties of the spin excitation\cite{Feng9404,Feng15}. According to
the previous discussions\cite{Feng9404,Feng15}, the spin propagators
$D^{(0)}({\bf k},\omega)$ and $D^{(0)}_{\rm z}({\bf k},\omega)$ in the MF
approximation can be calculated straightforwardly as,
\begin{subequations}\label{TWO-MFSGF}
\begin{eqnarray}
D^{(0)}({\bf k},\omega)&=&{B_{\bf k}\over 2\omega_{\bf k}}\left ({1\over \omega-
\omega_{\bf k}}-{1\over\omega+\omega_{\bf k}}\right ),\label{MFSGF}\\
D^{(0)}_{\rm z}({\bf k},\omega)&=& {B^{\rm (z)}_{\bf k}\over
2\omega^{\rm (z)}_{\bf k}}\left ({1\over\omega-\omega^{\rm (z)}_{\bf k}}
-{1\over\omega+\omega^{\rm (z)}_{\bf k}}\right ),~~~~
\label{MFSGFZ}
\end{eqnarray}
\end{subequations}
respectively, where the spin excitation spectra $\omega_{\bf k}$ and
$\omega^{\rm (z)}_{\bf k}$ in the MF approximation are respectively expressed as,
\begin{widetext}
\begin{subequations}\label{TWO-MFSES}
\begin{eqnarray}
\omega^{2}_{\bf k}&=&\alpha\lambda_{1}^{2}\left [{1\over 2}\epsilon\chi_{1}
\left (A_{11}-\gamma_{\bf k}\right)(\epsilon-\gamma_{\bf k})+\chi^{\rm z}_{1}
\left (A_{12}-\epsilon\gamma_{\bf k}\right )(1-\epsilon\gamma_{\bf k})\right ]
+\alpha\lambda_{2}^{2}\left [\left (\chi^{\rm z}_{2}\gamma_{\bf k}'
-{3\over 8}\chi_{2}\right )\gamma_{\bf k}'+A_{13}\right ]\nonumber\\
&+&\alpha\lambda_{1}\lambda_{2}\left [(\chi^{\rm z}_{1}+\chi^{\rm z}_{2})(A_{14}
-\epsilon\gamma_{\bf k})\gamma_{\bf k}'+{1\over 2}(\chi_{1}\gamma_{\bf k}'-C_{3})
(\epsilon-\gamma_{\bf k})-{1\over 2}\epsilon (C_{3}-\chi_{2}\gamma_{\bf k})
\right ], \label{MFSES}\\
(\omega^{\rm (z)}_{\bf k})^{2} &=&\alpha\epsilon\lambda^{2}_{1}(A_{15}
-\gamma_{\bf k})(1-\gamma_{\bf k})+\alpha\lambda^{2}_{2}A_{16}(1-\gamma_{\bf k}')
+\alpha\lambda_{1}\lambda_{2}[\epsilon C_{3}(\gamma_{\bf k}-1)
+(\chi_{2}\gamma_{\bf k}-\epsilon C_{3})(1 -\gamma_{\bf k}')],\label{MFSESZ}
\end{eqnarray}
\end{subequations}
\end{widetext}
while the corresponding weight functions $B_{{\bf{k}}}$ and $B_{{\rm z}{\bf k}}$ are
respectively given as,
\begin{subequations}
\begin{eqnarray}
B_{\bf k}&=&\lambda_{1}[2\chi^{\rm z}_{1}(\epsilon\gamma_{\bf k}-1)
+\chi_{1}(\gamma_{\bf k}-\epsilon)]\nonumber\\
&-&\lambda_{2}(2\chi^{\rm z}_{2}\gamma_{\bf k}'-\chi_{2}), ~~~~\\
B^{\rm (z)}_{\bf k} &=&\epsilon\chi_{1}\lambda_{1}(\gamma_{\bf k}-1)
-\chi_{2}\lambda_{2}(\gamma_{\bf k}'-1),~~~
\end{eqnarray}
\end{subequations}
with $\lambda_{1}=8J_{\rm eff}$, $\lambda_{2}=16\phi_{2}t'$,
$\epsilon=1+2t\phi_{1}/J_{\rm eff}$,
$\phi_{1}=\langle h^{\dagger}_{l\sigma}h_{l+\hat{\eta}\sigma}\rangle$,
$\phi_{2}=\langle h^{\dagger}_{l\sigma}h_{l+\hat{\tau}\sigma}\rangle$,
$A_{11}=[(1-\alpha)/(8\alpha)-\chi^{\rm z}_{1}/2+C_{1}]/\chi_{1}$,
$A_{12}=[(1-\alpha)/(16\alpha)-\epsilon\chi_{1}/8+C^{\rm z}_{1}]/\chi^{\rm z}_{1}$,
$A_{13}=[(1-\alpha)/(8\alpha)-\chi^{\rm z}_{2}/2+C_{2}]/2$,
$A_{14}=(\chi^{\rm z}_{1}+C_{3})/[\epsilon(\chi^{\rm z}_{1}+\chi^{\rm z}_{2})]$,
$A_{15}=\epsilon[(1-\alpha)/(8\alpha)+C_{1}]/\chi_{1}$,
$A_{16}=(1-\alpha)/(8\alpha)+C_{3}$, the spin correlation functions
$\chi^{\rm z}_{1}=\langle S_{l}^{\rm z}S_{l+\hat{\eta}}^{\rm z} \rangle$,
$\chi^{\rm z}_{2}=\langle S_{l}^{\rm z}S_{l+\hat{\tau}}^{\rm z} \rangle$,
$C_{1}=(1/16)\sum_{\hat{\eta},\hat{\eta'}}\langle S_{l+\hat{\eta}}^{+}
S_{l+\hat{\eta'}}^{-}\rangle$,
$C^{\rm z}_{1}=(1/16)\sum_{\hat{\eta},\hat{\eta'}}\langle S_{l+\hat{\eta}}^{z}
S_{l+\hat{\eta'}}^{z}\rangle$, $C_{2}=(1/16)\sum_{\hat{\tau},\hat{\tau'}}\langle S_{l+\hat{\tau}}^{+}S_{l+\hat{\tau'}}^{-}\rangle$,
$C_{3}=(1/4) \sum_{\hat{\tau}}\langle S_{l+\hat{\eta}}^{+}S_{l+\hat{\tau}}^{-}
\rangle$, and $C^{\rm z}_{3}=(1/4)\sum_{\hat{\tau}}\langle S_{l+\hat{\eta}}^{\rm z}
S_{l+\hat{\tau}}^{\rm z}\rangle$. In order to fulfill the sum rule of the spin
correlation function $\langle S^{+}_{l}S^{-}_{l}\rangle=1/2$ in the case without an
AFLRO, the crucial decoupling parameter $\alpha$ has been introduced in the above
calculation\cite{Feng9404,Feng15,Kondo72}, which can be regarded as the vertex
correction. At the half-filling, the $t$-$J$ model (\ref{CSS-tJ-model}) is reduced
as an isotropic Heisenberg model, and the degree of freedom is local spin only,
where $\epsilon=1$, $\lambda_{2}=0$, $\chi^{\rm z}_{1}=\chi_{1}/2$,
$C^{\rm z}_{1}=C_{1}/2$, and then the above spin excitation energy dispersion
$\omega_{\bf k}$ and the weight function $B_{\bf k}$ are respectively reduced as,
$\omega_{\bf k}=\lambda_{1}\sqrt{\alpha|\chi_{1}|(1-\gamma^{2}_{\bf k})}$ and
$B_{\bf k}=-2\lambda_{1}\chi_{1}(1-\gamma_{\bf k})$. This spin excitation energy
dispersion at the half-filling is the standard spin-wave dispersion. However, in the
doped regime without an AFLRO, although the form of the spin excitation energy
dispersion $\omega_{\bf k}$ in Eq. (\ref{MFSES}) is rather complicated, the essential
properties of the spin-wave nature are the same as that in the case of the
half-filling\cite{Feng9404,Feng15}.

\subsection{Charge-carrier normal and anomalous self-energies}
\label{Charge-carrier-self-energy}

In the framework of the kinetic-energy-driven superconductivity
\cite{Feng15,Feng0306,Feng12}, the spin-excitation mediates the charge-carrier
interaction directly from the kinetic energy of the $t$-$J$
model (\ref{CSS-tJ-model}), and then this charge-carrier interaction induces the
charge-carrier pairing state in the particle-particle channel and charge-carrier
pseudogap-state in the particle-hole channel, where the self-consistent equations
that are satisfied by the full charge-carrier diagonal and off-diagonal propagators
of the $t$-$J$ model (\ref{CSS-tJ-model}) are respectively derived as,
\begin{subequations}\label{CCSCES}
\begin{eqnarray}
g({\bf k},\omega)&=&g^{(0)}({\bf k},\omega)+g^{(0)}({\bf k},\omega)
[\Sigma^{(\rm h)}_{\rm ph}({\bf k},\omega)g({\bf k},\omega)\nonumber\\
&-&\Sigma^{(\rm h)}_{\rm pp}({\bf k},\omega)\Gamma^{\dagger}({\bf k},\omega)],~~~\\
\Gamma^{\dagger}({\bf k},\omega)&=& g^{(0)}({\bf k},-\omega)
[\Sigma^{(\rm h)}_{\rm ph}({\bf k},-\omega)\Gamma^{\dagger}({\bf k},\omega)
\nonumber\\
&+& \Sigma^{(\rm h)}_{\rm pp}({\bf k},\omega)g({\bf k},\omega)],~~~~
\end{eqnarray}
\end{subequations}
and then the full charge-carrier diagonal and off-diagonal propagators
$g({\bf k},\omega)$ and $\Gamma^{\dagger}({\bf k},\omega)$ are respectively expressed
explicitly as,
\begin{subequations}\label{CCDODGF}
\begin{eqnarray}
g({\bf k},\omega)&=&{1\over\omega-\xi_{\bf k}
-\Sigma^{(\rm h)}_{\rm tot}({\bf k},\omega)},~~~~~\label{CCDGF-1}\\
\Gamma^{\dagger}({\bf k},\omega)&=&{W_{\rm h}({\bf k},\omega)\over\omega
-\xi_{\bf k}-\Sigma^{(\rm h)}_{\rm tot}({\bf k},\omega)},
~~~~~~~\label{CCODGF-1}
\end{eqnarray}
\end{subequations}
where the charge-carrier total self-energy
$\Sigma^{(\rm h)}_{\rm tot}({\bf k},\omega)$ and the function
$W_{\rm h}({\bf k},\omega)$ are given as,
\begin{subequations}\label{CCTOTSE}
\begin{eqnarray}
\Sigma^{(\rm h)}_{\rm tot}({\bf k},\omega)&=&
\Sigma^{(\rm h)}_{\rm ph}({\bf k},\omega)
+{|\Sigma^{(\rm h)}_{\rm pp}({\bf k},\omega)|^{2}\over\omega+\xi_{\bf k}
+\Sigma^{(\rm h)}_{\rm ph}({\bf k},-\omega)},~~~~~~~~~~\\
W_{\rm h}({\bf k},\omega)&=&-{\Sigma^{(\rm h)}_{\rm pp}({\bf k},\omega)\over
\omega+\xi_{\bf k}+\Sigma^{(\rm h)}_{\rm ph}({\bf k},-\omega)},
\end{eqnarray}
\end{subequations}
respectively, with the charge-carrier normal self-energy
$\Sigma^{(\rm h)}_{\rm ph}({\bf k},\omega)$ in the particle-hole channel and the
charge-carrier anomalous self-energy $\Sigma^{(\rm h)}_{\rm pp}({\bf k},\omega)$
in the particle-particle channel, which have been evaluated in terms of the spin
bubble as\cite{Feng15,Feng0306,Feng12},
\begin{subequations}\label{CCSE}
\begin{eqnarray}
\Sigma^{({\rm h})}_{\rm ph}({\bf k},i\omega_{n})&=&{1\over N^{2}}
\sum_{{\bf p},{\bf p}'}\Lambda^{2}_{{\bf p}+{\bf p}'+{\bf k}}\nonumber\\
&\times& {1\over\beta}\sum_{ip_{m}}g({{\bf p}+{\bf k}},ip_{m}+i\omega_{n})
\Pi({\bf p},{\bf p}',ip_{m}), \nonumber\\
~~~~~~~~~ \label{CCNSE}\\
\Sigma^{({\rm h})}_{\rm pp}({\bf k},i\omega_{n})&=&{1\over N^{2}}
\sum_{{\bf p},{\bf p}'}\Lambda^{2}_{{\bf p}+{\bf p}'+{\bf k}}\nonumber\\
&\times& {1\over\beta}\sum_{ip_{m}}
\Gamma^{\dagger}({\bf p}+{\bf k},ip_{m}+i\omega_{n})\Pi({\bf p},{\bf p}',ip_{m}),
\nonumber\\
~~~~\label{CCANSE}
\end{eqnarray}
\end{subequations}
respectively, where $\omega_{n}$ and $p_{m}$ are the fermion and boson Matsubara
frequencies, respectively. The above charge-carrier self-energies in
Eq. (\ref{CCSE}) are induced by the coupling between charge and spin degrees of
freedom of the constrained electron directly from the kinetic energy of the $t$-$J$
model (\ref{CSS-tJ-model}) in the fermion-spin representation\cite{Feng15}, where
the bare vertex function $\Lambda_{\bf k}=4t\gamma_{\bf k}-4t'\gamma_{\bf{k}}'$ is
originated from the derivation of these charge-carrier self-energies in the random
phase approximation. The spin bubble $\Pi({\bf p},{\bf p}',ip_{m})$ is obtained in
terms of the spin propagator $D^{(0)}({\bf k},\omega)$ in Eq. (\ref{MFSGF}) as,
\begin{equation}
\Pi({\bf p},{\bf p}',ip_{m})={1\over\beta}\sum_{ip_{m}'}D^{(0)}({\bf p}',ip_{m}')
D^{(0)}({\bf p'+p},ip_{m}'+ip_{m}).
\end{equation}
The charge-carrier anomalous self-energy $\Sigma^{(\rm h)}_{\rm pp}({\bf k},\omega)$
is identified as the momentum and energy dependence of the charge-carrier pair gap,
i.e., $\Sigma^{({\rm h})}_{\rm pp}({\bf k},\omega)
=\bar{\Delta}^{\rm (h)}_{\bf k}(\omega)$, while the charge-carrier normal
self-energy $\Sigma^{(\rm h)}_{\rm ph}({\bf k},\omega)$ describes the momentum and
energy dependence of the charge-carrier single-particle coherence. Although
$\Sigma^{({\rm h})}_{\rm pp}({\bf k},\omega)$ is an even function of energy,
$\Sigma^{({\rm h})}_{\rm ph}({\bf k},\omega)$ is not. In this case,
$\Sigma^{({\rm h})}_{\rm ph}({\bf k},\omega)$ can be divided into its symmetric and
antisymmetric parts as: $\Sigma^{({\rm h})}_{\rm ph}({\bf k},\omega)
=\Sigma^{({\rm h})}_{\rm phe}({\bf k},\omega)
+\omega\Sigma^{({\rm h})}_{\rm pho}({\bf k},\omega)$, and then both
$\Sigma^{({\rm h})}_{\rm phe}({\bf k},\omega)$ and
$\Sigma^{({\rm h})}_{\rm pho}({\bf k},\omega)$ are an even function of energy.
Furthermore, this antisymmetric part $\Sigma^{({\rm h})}_{\rm pho}({\bf k},\omega)$
is correlated directly with the momentum and energy dependence of the charge-carrier
single-particle coherent weight as: $Z^{{\rm (h)}-1}_{\rm F}({\bf k},\omega)
=1-{\rm Re}\Sigma^{(\rm h)}_{\rm pho}({\bf k},\omega)$. In this paper, we focus
mainly on the anomalous properties of the low-energy electronic structure, and then
$\bar{\Delta}^{\rm (h)}_{\bf k}(\omega)$ and $Z^{{\rm (h)}}_{\rm F}({\bf k},\omega)$
can be discussed in the static limit, i.e.,
\begin{subequations}
\begin{eqnarray}
\bar{\Delta}^{\rm (h)}_{\bf k}&=&\bar{\Delta}_{\rm h}\gamma^{\rm (d)}_{\bf k},
\label{CCPGF}\\
{1\over Z^{\rm (h)}_{\rm F}({\bf k})}&=&1
-{\rm Re}\Sigma^{(\rm h)}_{\rm pho}({\bf k},\omega)\mid_{\omega=0},
\label{CCQCWW}
\end{eqnarray}
\end{subequations}
with the charge-carrier pair gap parameter $\bar{\Delta}_{\rm h}$. Although
$Z^{\rm (h)}_{\rm F}({\bf k})$ still is a function of momentum, the momentum
dependence is unimportant in a qualitative discussion. Following the ARPES
experiments\cite{DLFeng00,Ding01}, the momentum
${\bf k}$ in $Z^{\rm (h)}_{\rm F}({\bf k})$ can be chosen as,
\begin{eqnarray}\label{CCQCW}
Z^{\rm (h)}_{\rm F}=Z^{\rm (h)}_{\rm F}({\bf k})\mid_{{\bf k}=[\pi,0]}.
\end{eqnarray}
With the above static-limit approximation, the renormalized charge-carrier
diagonal and off-diagonal propagators now can be obtained from Eq. (\ref{CCDODGF})
as,
\begin{subequations}\label{BCSHGF}
\begin{eqnarray}
g^{\rm (RMF)}({\bf k},\omega)&=&Z^{\rm (h)}_{\rm F}\left ( {U^{2}_{\rm h}({\bf k})
\over\omega-E^{\rm (h)}_{\bf k}}+{V^{2}_{\rm h}({\bf k})\over\omega
+E^{\rm (h)}_{\bf k}}\right ),\label{BCSHDGF}\\
\Gamma^{{\rm (RMF)}\dagger}({\bf k},\omega)&=&-Z^{\rm (h)}_{\rm F}
{\bar{\Delta}^{\rm (h)}_{{\rm Z}{\bf k}}\over 2E^{\rm (h)}_{\bf k}}
\left ( {1\over \omega-E^{\rm (h)}_{\bf k}}-{1\over\omega+E^{\rm (h)}_{\bf k}}
\right ),~~\nonumber\\
~~~\label{BCSHODGF}
\end{eqnarray}
\end{subequations}
with the charge-carrier quasiparticle energy dispersion
$E^{\rm (h)}_{\bf k}=\sqrt{\bar{\xi}^{2}_{\bf k}
+\mid\bar{\Delta}^{\rm (h)}_{{\rm Z}{\bf k}}\mid^{2}}$, the renormalized MF
charge-carrier energy dispersion
$\bar{\xi}_{{\bf k}}=Z^{\rm (h)}_{\rm F}\xi_{\bf k}$, the renormalized
charge-carrier pair gap $\bar{\Delta}^{\rm (h)}_{{\rm Z}{\bf k}}
=Z^{\rm (h)}_{\rm F}\bar{\Delta}^{\rm (h)}_{\bf k}$, and the charge-carrier
quasiparticle coherence factors,
\begin{subequations}\label{BCSCF}
\begin{eqnarray}
U^{2}_{\rm h}({\bf k})&=&{1\over 2}\left (1+{\bar{\xi}_{{\bf k}}
\over E^{\rm (h)}_{\bf k}}\right ),\\
V^{2}_{\rm h}({\bf k})&=&{1\over 2}\left (1-{\bar{\xi}_{{\bf k}}
\over E^{\rm (h)}_{\bf k}}\right ),
\end{eqnarray}
\end{subequations}
which satisfy the constraint $U^{2}_{\rm h}({\bf k})+V^{2}_{\rm h}({\bf k})=1$ for
any momentum ${\bf k}$.

Substituting the renormalized charge-carrier diagonal and off-diagonal propagators
in Eq. (\ref{BCSHGF}) and spin propagator in Eq. (\ref{MFSGF}) into
Eq. (\ref{CCSE}), the charge-carrier normal self-energy
$\Sigma^{({\rm h})}_{\rm ph}({\bf k},\omega)$ and the anomalous self-energy
$\Sigma^{({\rm h})}_{\rm pp}({\bf k},\omega)$ can be respectively derived as,
\begin{widetext}
\begin{subequations}\label{SE1}
\begin{eqnarray}
{\Sigma}^{\rm(h)}_{\rm ph}({\bf{k}},\omega)&=&{Z^{\rm (h)}_{\rm F}\over N^{2}}
\sum_{{\bf{p}}{\bf{p}'}{\nu}}(-1)^{\nu+1}
\Omega_{{\bf{p}}{\bf{p}'}{\bf{k}}}\left [ U^{2}_{\rm{h}}({\bf p}+{\bf k})\left (
{F^{(\rm{h})}_{1\nu}({\bf p},{\bf p}',{\bf k})\over\omega
+\omega^{(\nu)}_{{\bf p}{\bf p}'}-E^{\rm (h)}_{{\bf p}+{\bf k}}}
- {F^{(\rm{h})}_{2\nu}({\bf p},{\bf p}',{\bf k})\over\omega
-\omega^{(\nu)}_{{\bf p}{\bf p}'}-E^{\rm (h)}_{{\bf p}+{\bf k}}}\right )\right.
\nonumber\\
&+&\left. V^{2}_{{\rm h}}({\bf p}+{\bf k})\left (
{F^{(\rm{h})}_{1\nu}({\bf p},{\bf p}',{\bf k})\over\omega-
\omega^{(\nu)}_{{\bf p}{\bf p}'}+E^{\rm (h)}_{{\bf p}+{\bf k}}}
-  {F^{(\rm{h})}_{2\nu}({\bf p},{\bf p}',{\bf k})\over\omega
+\omega^{(\nu)}_{{\bf p}{\bf p}'}+E^{\rm (h)}_{{\bf p}+{\bf k}}}\right ) \right ],
\label{NSE1}~~~~\\
{\Sigma}^{\rm(h)}_{{\rm pp}}({\bf{k}},\omega)&=&{Z^{\rm (h)}_{\rm F}\over N^{2}}
\sum_{{\bf{p}}{\bf{p}'}{\nu}}(-1)^{\nu}
\Omega_{{\bf{p}}{\bf{p}'}{\bf{k}}}{\bar{\Delta}^{\rm (h)}_{{\rm Z}{\bf p}+{\bf k}}
\over 2E^{\rm (h)}_{{\bf p}+{\bf k}}}\left [ \left (
{F^{(\rm{h})}_{1\nu}({\bf p},{\bf p}',{\bf k})\over\omega
+\omega^{(\nu)}_{{\bf p}{\bf p}'}-E^{\rm (h)}_{{\bf p}+{\bf k}}}
-{F^{(\rm{h})}_{2\nu}({\bf p},{\bf p}',{\bf k})\over\omega
-\omega^{(\nu)}_{{\bf p}{\bf p}'}-E^{\rm (h)}_{{\bf p}+{\bf k}}}\right )\right.
\nonumber\\
&-&\left. \left ( {F^{(\rm{h})}_{1\nu}({\bf p},{\bf p}',{\bf k})\over\omega
-\omega^{(\nu)}_{{\bf p}{\bf p}'}+E^{\rm (h)}_{{\bf p}+{\bf k}}}
- {F^{(\rm{h})}_{2\nu}({\bf p},{\bf p}',{\bf k})\over\omega
+\omega^{(\nu)}_{{\bf p}{\bf p}'}+E^{\rm (h)}_{{\bf p}+{\bf k}}}\right ) \right ],
\label{ANSE1}
\end{eqnarray}
\end{subequations}
with $\nu=1,2$, the kernel function ${\Omega}_{{\bf{p}}{\bf{p}'}{\bf{k}}}
=[\Lambda_{{\bf{p}}+{\bf{p}'}+{\bf{k}}}]^{2}B_{\bf{p}'}B_{\bf{p}+\bf{p}'}
/(4\omega_{{\bf p}'}\omega_{{\bf p}+{\bf p}'})$, $\omega^{(\nu)}_{{\bf p}{\bf p}'}
={\omega}_{\bf{p}+\bf{p}'}-(-1)^{\nu}{\omega}_{\bf{p}'}$, and the functions,
\begin{subequations}
\begin{eqnarray}
F^{(\rm{h})}_{1\nu}({\bf p},{\bf p}',{\bf k})&=&n_{\rm{F}}
[E^{\rm (h)}_{{\bf p}+{\bf k}}]\{1+n_{\rm B}(\omega_{\bf{p}'+\bf{p}})
+ n_{\rm B}[(-1)^{\nu+1}\omega_{\bf{p}'}]\}
+n_{\rm B}(\omega_{\bf{p}'+\bf{p}})n_{\rm B}[(-1)^{\nu+1}\omega_{\bf{p}'}],\\
F^{(\rm{h})}_{2\nu}({\bf{p}},{\bf{p}'},{\bf{k}})&=&\{1-n_{\rm{F}}
[E^{\rm (h)}_{{\bf p}+{\bf k}}]\}\{1+n_{\rm B}(\omega_{\bf{p}'+\bf{p}})
+n_{\rm B}[(-1)^{\nu+1}\omega_{\bf{p}'}]\}
+n_{B}(\omega_{\bf{p}'+\bf{p}})n_{B}[(-1)^{\nu+1}\omega_{\bf{p}'}],~~~~~~~~
\end{eqnarray}
\end{subequations}
where $n_{B}(\omega)$ and $n_{F}(\omega)$ are the boson and fermion distribution
functions, respectively.

\subsection{Self-consistent equations for determination of charge-carrier order
parameters}\label{CSCE}

The above charge-carrier single-particle coherent weight $Z^{\rm (h)}_{\rm F}$ and
the charge-carrier pair gap parameter $\bar{\Delta}_{\rm h}$ respectively satisfy
following two self-consistent equations,
\begin{subequations}\label{SCE1}
\begin{eqnarray}
{1\over Z^{\rm (h)}_{\rm F}} &=& 1+{Z^{\rm (h)}_{\rm F}\over N^{2}}
\sum_{{\bf{p}}{\bf{p}'}{\nu}}(-1)^{\nu+1}
{\Omega}_{{\bf{p}}{\bf{p}'}{{\bf k}_{\rm A}}}
\left ( {F^{(\rm{h})}_{1\nu}({\bf{p}},{\bf{p}}',{\bf k}_{\rm A})
\over [\omega^{(\nu)}_{{\bf{p}}{\bf{p}}'}-E^{\rm (h)}_{\bf{p}+{\bf k}_{\rm A}}]^{2}}
+ {F^{(\rm{h})}_{2\nu}({\bf{p}},{\bf{p}}',{\bf k}_{\rm A})
\over [\omega^{(\nu)}_{{\bf{p}}{\bf{p}}'}
+E^{\rm (h)}_{\bf{p}+\bf{k}_{\rm A}}]^{2}} \right ), \\
1 &=& {4[Z^{\rm (h)}_{\rm F}]^{2}\over N^{3}}
\sum_{{\bf{p}}{\bf{p}'}{\bf{k}}{\nu}}(-1)^{\nu}\Omega_{{\bf{p}}{\bf{p}'}{\bf{k}}}
{\gamma^{\rm (d)}_{\bf k}\gamma^{\rm (d)}_{{\bf p}+{\bf k}}
\over E^{\rm (h)}_{{\bf p}+{\bf k}}}
\left( {F^{(\rm{h})}_{1\nu}({\bf{p}},{\bf{p}}',{\bf{k}})
\over \omega^{(\nu)}_{{\bf{p}}{\bf{p}}'}
-E^{\rm (h)}_{\bf{p}+\bf{k}}}-{F^{(\rm{h})}_{2\nu}({\bf{p}},{\bf{p}}',{\bf{k}})
\over\omega^{(\nu)}_{{\bf{p}}{\bf{p}}'}+E^{\rm (h)}_{\bf{p}+\bf{k}}} \right ),
\end{eqnarray}
\end{subequations}
\end{widetext}
where ${\bf k}_{\rm A}=[\pi,0]$. The above two self-consistent equations must be
solved simultaneously with following self-consistent equations,
\begin{subequations}\label{SCE2}
\begin{eqnarray}
\phi_{1}&=&{Z^{\rm (h)}_{\rm F}\over 2N}\sum_{{\bf k}}\gamma_{{\bf k}}
\left (1-{\bar{\xi_{{\bf k}}}\over E^{\rm (h)}_{\bf k}}{\rm tanh}
[{1\over 2}\beta E^{\rm (h)}_{\bf k}]\right ),~~~~~~~~~\\
\phi_{2}&=&{Z^{\rm (h)}_{\rm F}\over 2N}\sum_{{\bf k}}\gamma_{{\bf k}}'
\left (1-{\bar{\xi_{{\bf k}}}\over E^{\rm (h)}_{\bf k}}{\rm tanh}
[{1\over 2}\beta E^{\rm (h)}_{\bf k}]\right ),~~~~~\\
\delta &=& {Z^{\rm (h)}_{\rm F}\over 2N}\sum_{{\bf k}}
\left (1-{\bar{\xi_{{\bf k}}}\over E^{\rm (h)}_{\bf k}}{\rm tanh}
[{1\over 2}\beta E^{\rm (h)}_{\bf k}] \right ),~~~~\\
\chi_{1}&=&{1\over N}\sum_{{\bf k}}\gamma_{{\bf k}} {B_{{\bf k}}\over
2\omega_{{\bf k}}}{\rm coth} [{1\over 2}\beta\omega_{{\bf k}}], \\
\chi_{2}&=&{1\over N}\sum_{{\bf k}}\gamma_{{\bf k}}'{B_{{\bf k}}\over
2\omega_{{\bf k}}}{\rm coth} [{1\over 2}\beta\omega_{{\bf k}}],\\
C_{1}&=&{1\over N}\sum_{{\bf k}}\gamma^{2}_{{\bf k}} {B_{{\bf k}}\over
2\omega_{{\bf k}}}{\rm coth}[{1\over 2}\beta\omega_{{\bf k}}],\\
C_{2}&=&{1\over N}\sum_{{\bf k}}\gamma'^{2}_{{\bf k}} {B_{{\bf k}}\over
2\omega_{{\bf k}}}{\rm coth}  [{1\over 2}\beta\omega_{{\bf k}}], \\
C_{3}&=&{1\over N}\sum_{{\bf k}}\gamma_{{\bf k}}\gamma_{{\bf k}}'{B_{{\bf k}}
\over 2\omega_{{\bf k}}}{\rm coth}[{1\over 2}\beta\omega_{{\bf k}}],\\
{1\over 2} &=&{1\over N}\sum_{{\bf k}}{B_{{\bf k}} \over
2\omega_{{\bf k}}}{\rm coth} [{1\over 2}\beta\omega_{{\bf k}}],\label{SCE2i}\\
\chi^{\rm z}_{1}&=&{1\over N}\sum_{{\bf k}}\gamma_{{\bf k}} {B^{\rm (z)}_{\bf k}
\over 2\omega^{\rm (z)}_{\bf k}}{\rm coth}
[{1\over 2}\beta\omega^{\rm (z)}_{\bf k}],\\
\chi^{\rm z}_{2}&=& {1\over N}\sum_{{\bf k}}\gamma_{{\bf k}}'{B^{\rm (z)}_{\bf k}
\over 2\omega^{\rm (z)}_{\bf k}}{\rm coth}
[{1\over 2}\beta\omega^{\rm (z)}_{\bf k}],
\end{eqnarray}
\begin{eqnarray}
C^{\rm z}_{1}&= &{1\over N}\sum_{{\bf k}}\gamma^{2}_{{\bf k}}{B^{\rm (z)}_{\bf k}
\over 2\omega^{\rm (z)}_{\bf k}}{\rm coth}
[{1\over 2}\beta\omega^{\rm (z)}_{\bf k}], \\
C^{\rm z}_{3}&=&{1\over N}\sum_{{\bf k}}\gamma_{{\bf k}}
\gamma_{{\bf k}}'{B^{\rm (z)}_{\bf k}\over 2\omega^{\rm (z)}_{\bf k}}
{\rm coth} [{1\over 2}\beta\omega^{\rm (z)}_{\bf k}],
\end{eqnarray}
\end{subequations}
and then all the above order parameters, charge-carrier's particle-hole
parameters, spin correlation functions, decoupling parameter $\alpha$,
and charge-carrier chemical potential $\mu_{\rm h}$ are determined by the
self-consistent calculation {\it without using any adjustable parameters}.

The above self-consistent equations (\ref{SCE1}) and (\ref{SCE2}) have been
calculated at different doping concentrations\cite{Feng15,Feng0306,Feng12}, and
the result shows that the charge-carrier pair gap parameter $\bar{\Delta}_{\rm h}$
exhibits a dome-like shape doping dependence, i.e., with the increase of doping,
$\bar{\Delta}_{\rm h}$ is raised gradually in the underdoped regime, and reaches
its maximum at around the optimal doping $\delta\approx 0.15$, then with a further
increase of doping, $\bar{\Delta}_{\rm h}$ turns into a monotonically decreases in
the overdoped regime. It should be emphasized that the above parameters determined
by the self-consistent equation (\ref{SCE2}), including the charge-carrier's
particle-hole parameters
$\phi_{1}=\langle h^{\dagger}_{l\sigma}h_{l+\hat{\eta}\sigma} \rangle$ and
$\phi_{2}=\langle h^{\dagger}_{l\sigma}h_{l+\hat{\tau}\sigma}\rangle$, the spin
correlation functions $\chi_{1}=\langle S_{l}^{+}S_{l+\hat{\eta}}^{-}\rangle$,
$\chi_{2}=\langle S_{l}^{+}S_{l+\hat{\tau}}^{-}\rangle$,
$\chi^{\rm z}_{1}=\langle S_{l}^{\rm z}S_{l+\hat{\eta}}^{\rm z}\rangle$,
$\chi^{\rm z}_{2}=\langle S_{l}^{\rm z}S_{l+\hat{\tau}}^{\rm z} \rangle$,
$C_{1}=(1/Z^{2})\sum_{\hat{\eta},\hat{\eta'}}\langle S_{l+\hat{\eta}}^{+}
S_{l+\hat{\eta'}}^{-}\rangle$, $C^{\rm z}_{1}=(1/Z^{2})\sum_{\hat{\eta},\hat{\eta'}}
\langle S_{l+\hat{\eta}}^{z}S_{l+\hat{\eta'}}^{z}\rangle$,
$C_{2}=(1/Z^{2})\sum_{\hat{\tau},\hat{\tau'}}\langle S_{l+\hat{\tau}}^{+}
S_{l+\hat{\tau'}}^{-}\rangle$, $C_{3}=(1/Z) \sum_{\hat{\tau}}\langle
S_{l+\hat{\eta}}^{+}S_{l+\hat{\tau}}^{-}\rangle$,
$C^{\rm z}_{3}=(1/Z)\sum_{\hat{\tau}}\langle S_{l+\hat{\eta}}^{\rm z}
S_{l+\hat{\tau}}^{\rm z}\rangle$, and the decoupling parameter $\alpha$ in the
Kondo-Yamaji decoupling approximation\cite{Kondo72} for the derivation of the
spin propagators in Eq. (\ref{TWO-MFSGF}), are essential MF parameters for
the description of the charge-carrier and spin parts of the MF $t$-$J$ model in
Eq. (\ref{MF-t-J-model}), and therefore the magnitude of these parameters are finite
both in the pseudogap phase (then the normal-state) and SC-state.

\subsection{Charge-carrier pair transition temperature}\label{CCPTT}

The charge-carrier pair gap $\bar{\Delta}_{\rm h}$ is strongly temperature
dependent, and vanishes at the charge-carrier pair transition temperature
$T^{\rm (pair)}_{\rm c}$. This $T^{\rm (pair)}_{\rm c}$ has been also evaluated
self-consistently at different doping concentrations\cite{Feng15,Feng0306,Feng12},
and the calculated result shows that $T^{\rm (pair)}_{\rm c}$ as a function of
doping presents a similar behavior of $\bar{\Delta}_{\rm h}$, i.e., the maximal
$T^{\rm (pair)}_{\rm c}$ emerges at around the optimal doping $\delta\approx 0.15$,
and then decreases in both the underdoped and the overdoped regimes. Moreover,
we\cite{Feng15a} have shown that the magnitude of $T^{\rm (pair)}_{\rm c}$ is
exactly the same as the corresponding one of the SC transition temperature $T_{\rm c}$
derived from the electron pairing state at the condition of the electron pair gap
parameter $\bar{\Delta}=0$, and we will return to this discussion of
$T^{\rm (pair)}_{\rm c}$ towards subsection \ref{EPTT} of this Appendix.

\subsection{Pseudogap in charge channel}\label{charge-pseudogap}

To explore the nature of the pseudogap in the charge channel, we rewrite the
charge-carrier normal self-energy in Eq. (\ref{NSE1}) as,
\begin{eqnarray}\label{self-energy-2}
\Sigma^{({\rm h})}_{\rm ph}(\bf k,\omega)&\approx&
{[2\bar{\Delta}^{({\rm h})}_{\rm pg}({\bf k})]^{2}\over\omega-\xi_{0{\bf k}}},
\end{eqnarray}
where $\xi_{0{\bf k}}=L^{({\rm h})}_{2}({\bf k})/L^{({\rm h})}_{1}({\bf k})$ is the
energy spectrum of $\Sigma^{({\rm h})}_{\rm ph}(\bf k,\omega)$, while
$\bar{\Delta}^{({\rm h})}_{\rm pg}({\bf k})=L^{({\rm h})}_{2}({\bf k})
/\sqrt{4L^{({\rm h})}_{1}({\bf k})}$ is refereed to as the momentum dependence of
the charge-carrier pseudogap, since it plays a role of the suppression of the
charge-carrier density of states. The functions
$L^{({\rm h})}_{1}({\bf k})=-\Sigma^{({\rm h})}_{\rm pho}({\bf k},\omega=0)$ and
$L^{({\rm h})}_{2}({\bf k})=-\Sigma^{({\rm h})}_{\rm ph}({\bf k},\omega=0)$, while
$\Sigma^{({\rm h})}_{\rm ph}({\bf k},\omega=0)$ and the antisymmetric part
$\Sigma^{({\rm h})}_{\rm pho}({\bf k},\omega)$ of the charge-carrier normal
self-energy are calculated directly from
$\Sigma^{({\rm h})}_{\rm ph}({\bf k},\omega)$ in Eq. (\ref{NSE1}). In this case,
it is straightforward to obtain the charge-carrier pseudogap parameter as
$\bar{\Delta}^{({\rm h})}_{\rm pg}=(1/N)\sum_{\bf k}
\bar{\Delta}^{({\rm h})}_{\rm pg}({\bf k})$.

In the previous discussions\cite{Feng15,Feng12}, the doping dependence of the
charge-carrier pseudogap $\bar{\Delta}^{({\rm h})}_{\rm pg}$ has been discussed in
the detail, where the quite large $\bar{\Delta}^{({\rm h})}_{\rm pg}$ appears in
the underdoped regime, then it weakens as the doping is increased, and eventually
disappears together with $\bar{\Delta}_{\rm h}$ in the heavily overdoped region.
In particular, at a given doping, this $\bar{\Delta}^{({\rm h})}_{\rm pg}$ is
identified as a crossover with a charge-carrier pseudogap crossover temperature
$T^{*}_{\rm h}$, and then in conformity with the doping dependence of
$\bar{\Delta}^{({\rm h})}_{\rm pg}$, $T^{*}_{\rm h}$ is quite high at the
underdoped regime, and then it smoothly decreases with the increase of doping,
eventually terminating together with $T^{\rm (pair)}_{\rm c}$ at the heavily
overdoped region\cite{Feng15,Feng12}. More importantly, as the case of the electron
pairing state, the normal-state pseudogap state originated from the charge-carrier
pseudogap state is also due to the charge-spin recombination\cite{Feng15a}, and we
will return to the discussion of the normal-state pseudogap towards subsection
\ref{FCSRS}.

\subsection{Full charge-spin recombination}\label{FCSRS}

For the investigation of the low-energy electronic structure of the cuprate
superconductors in the SC-state, we need to derive the full electron diagonal and
off-diagonal propagators $G({\bf k},\omega)$ and $\Im^{\dagger}({\bf k},\omega)$ in
Eq. (\ref{EDODGF}) of the main text. In the previous studies\cite{Feng15a}, the
theory of the full charge-spin recombination has been established, where a charge
carrier and a localized spin are fully recombined into a constrained electron. In
particular, within this framework of the full charge-spin recombination\cite{Feng15a},
it has been realized that the coupling form between the electrons and spin
excitations is the same as the form between the charge carriers and spin excitations,
which implies that the form of the self-consistent equations satisfied by the full
electron diagonal and off-diagonal propagators is the same as that satisfied by the
full charge-carrier diagonal and off-diagonal propagators. Following these previous
discussions\cite{Feng15a}, we can perform a full charge-spin recombination in
which the full charge-carrier diagonal and off-diagonal propagators
$g({\bf k},\omega)$ and $\Gamma^{\dagger}({\bf k},\omega)$ in Eq. (\ref{CCSCES}) are
respectively replaced by the full electron diagonal and off-diagonal propagators
$G({\bf k},\omega)$ and $\Im^{\dagger}({\bf k},\omega)$, and then the self-consistent
equations satisfied by the full electron diagonal and off-diagonal propagators of the
$t$-$J$ model (\ref{tjham}) in the SC-state can be obtained straightforwardly as,
\begin{subequations}\label{ESCES}
\begin{eqnarray}
G({\bf k},\omega)&=&G^{(0)}({\bf k},\omega)+G^{(0)}({\bf k},\omega)
[\Sigma_{\rm ph}({\bf k},\omega)G({\bf k},\omega)\nonumber\\
&-& \Sigma_{\rm pp}({\bf k},\omega)\Im^{\dagger}({\bf k},\omega)],~~~~~~~~~~
\label{EDGF} \\
\Im^{\dagger}({\bf k},\omega)&=&G^{(0)}({\bf k},-\omega)
[\Sigma_{\rm ph}({\bf k},-\omega)\Im^{\dagger}({\bf k},\omega)\nonumber\\
&+& \Sigma_{\rm pp}({\bf k},\omega)G({\bf k},\omega)], \label{EODGF}
\end{eqnarray}
\end{subequations}
with the electron diagonal propagator $G^{(0)}({\bf k},\omega)$ of the $t$-$J$
model (\ref{tjham}) in the tight-binding approximation,
\begin{eqnarray}\label{MFEGF}
G^{(0)}({\bf k},\omega)&=&{1\over \omega-\varepsilon_{\bf k}},
\end{eqnarray}
and then the full electron diagonal and off-diagonal propagators can be expressed
explicitly as,
\begin{subequations}\label{EDODGF-1}
\begin{eqnarray}
G({\bf k},\omega)&=&{1\over\omega-\varepsilon_{\bf k}
-\Sigma_{\rm tot}({\bf k},\omega)},~~~~~\label{EDGF-1}\\
\Im^{\dagger}({\bf k},\omega)&=&{W({\bf k},\omega)\over\omega
-\varepsilon_{\bf k}-\Sigma_{\rm tot}({\bf k},\omega)},
~~~~~~~\label{EODGF-1}
\end{eqnarray}
\end{subequations}
as quoted in Eq. (\ref{EDODGF}) of the main text, while the electron normal
self-energy $\Sigma_{\rm ph}({\bf k},\omega)$ in the particle-hole channel and
the electron anomalous self-energy $\Sigma_{\rm pp}({\bf k},\omega)$ in the
particle-particle channel are respectively obtained from the corresponding parts of
the charge-carrier normal self-energy $\Sigma^{(\rm h)}_{\rm ph}({\bf k},\omega)$
in Eq. (\ref{CCNSE}) and the charge-carrier anomalous self-energy
$\Sigma^{(\rm h)}_{\rm pp}({\bf k},\omega)$ in Eq. (\ref{CCANSE}) by the replacement
of the full charge-carrier diagonal and off-diagonal propagators $g({\bf k},\omega)$
and $\Gamma^{\dagger}({\bf k},\omega)$ with the corresponding full electron diagonal
and off-diagonal propagators $G({\bf k},\omega)$ and $\Im^{\dagger}({\bf k},\omega)$.
However, in the previous discussions \cite{Feng16,Gao18,Gao19}, the calculations of
the electron normal and anomalous self-energies suffer from ignoring the vertex
correction. For a better description of the anomalous properties of the cuprate
superconductors, the electron normal and anomalous self-energies have been re-derived
in the recent studies\cite{Zeng25} by taking into account the vertex correction,
and the calculated results show that the vertex correction functions are energy and
momentum dependent. However, in this paper, as a qualitative discussion of the
anomalous properties of the electronic state in the case for the low-energy close to
EFS, the vertex corrections for the electron normal and anomalous self-energies can
be discussed in the static limit, and then the electron normal and anomalous
self-energies are respectively calculated as,
\begin{subequations}\label{ESE}
\begin{eqnarray}
\Sigma_{\rm ph}({\bf k},i\omega_{n})&=&{1\over N^{2}}\sum_{{\bf p},{\bf p}'}
[V_{\rm ph}\Lambda_{{\bf p}+{\bf p}'+{\bf k}}]^{2}\nonumber\\
&\times& {1\over\beta}\sum_{ip_{m}}G({{\bf p}+{\bf k}},ip_{m}+i\omega_{n})
\Pi({\bf p},{\bf p}',ip_{m}),\nonumber\\
~~~~~~~~~~\\
\Sigma_{\rm pp}({\bf k},i\omega_{n})&=&{1\over N^{2}}\sum_{{\bf p},{\bf p}'}
[V_{\rm pp}\Lambda_{{\bf p}+{\bf p}'+{\bf k}}]^{2}\nonumber\\
&\times& {1\over \beta}\sum_{ip_{m}}\Im^{\dagger}({\bf p}+{\bf k},ip_{m}+i\omega_{n})
\Pi({\bf p},{\bf p}',ip_{m}),\nonumber\\
~~~~~
\end{eqnarray}
\end{subequations}
as quoted in Eq. (\ref{E-N-AN-SE}) of the main text.

The calculation process of the electron normal and anomalous self-energies is the
same as that of the charge-carrier normal and anomalous self-energies in subsection
\ref{Charge-carrier-self-energy}, i.e., the electron normal self-energy
$\Sigma_{\rm ph}({\bf k},\omega)$ is separated into its symmetric and antisymmetric
parts as: $\Sigma_{\rm ph}({\bf k},\omega)=\Sigma_{\rm phe}({\bf k},\omega)
+\omega\Sigma_{\rm pho}({\bf k},\omega)$, and then both the symmetric part
$\Sigma_{\rm phe}({\bf k},\omega)$ and antisymmetric part
$\Sigma_{\rm pho}({\bf k},\omega)$ are an even function of energy. However, this
antisymmetric part $\Sigma_{\rm pho}({\bf k},\omega)$ is identified as the
single-particle coherent weight:
$Z^{-1}_{\rm F}({\bf k},\omega)=1-{\rm Re}\Sigma_{\rm pho}({\bf k},\omega)$.
In an interacting system, everything happens near EFS. In the case of the discussions
of the low-energy electronic structure, the electron pair gap and single-particle
coherent weight can be discussed in the static-limit as,
\begin{subequations}\label{EPGF-EQCW}
\begin{eqnarray}
\bar{\Delta}_{\bf k}&=&\bar{\Delta}\gamma^{\rm (d)}_{\bf k},\label{EPGF}\\
{1\over Z_{\rm F}}&=&1-{\rm Re}\Sigma_{\rm pho}({\bf k},0)\mid_{{\bf k}=[\pi,0]}, ~~~
\label{EQCW}
\end{eqnarray}
\end{subequations}
where the wave vector ${\bf k}$ in $Z_{\rm F}({\bf k})$ has been chosen as
${\bf k}=[\pi,0]$ just as it has been done in the ARPES experiments
\cite{Ding01,DLFeng00}.

With the help of the above the electron pair gap $\bar{\Delta}_{\bf k}$ and
single-particle coherent weight $Z_{\rm F}$ in Eq. (\ref{EPGF-EQCW}), the
renormalized electron diagonal and off-diagonal propagators now can be calculated
from Eq. (\ref{ESCES}) as,
\begin{subequations}\label{MF-EGFS}
\begin{eqnarray}
G^{\rm (RMF)}({\bf k},\omega)&=&Z_{\rm F}\left ({U^{2}_{\bf k}\over
\omega-E_{\bf k}}+{V^{2}_{\bf k}\over\omega +E_{\bf k}} \right ), ~~~~~~\\
\Im^{{\rm (RMF)}\dagger}({\bf k},\omega)&=&-Z_{\rm F}
{\bar{\Delta}_{{\rm Z}{\bf k}}\over 2E_{\bf k}}\left ({1\over\omega
-E_{\bf k}}-{1\over\omega+E_{\bf k}}\right ),\nonumber\\
~~~
\end{eqnarray}
\end{subequations}
where the renormalized electron energy dispersion
$\bar{\varepsilon}_{\bf k}=Z_{\rm F}\varepsilon_{\bf k}$, the renormalized electron
pair gap $\bar{\Delta}_{{\rm Z}{\bf k}}=Z_{\rm F}\bar{\Delta}_{\bf k}$, the SC
quasiparticle energy dispersion $E_{\bf k}=\sqrt {\bar{\varepsilon}^{2}_{\bf k}
+|\bar{\Delta}_{{\rm Z}{\bf k}}|^{2}}$, and the quasiparticle coherence factors
\begin{subequations}\label{EBCSCF}
\begin{eqnarray}
U^{2}_{\bf k}&=&{1\over 2}\left ( 1+{\bar{\varepsilon}_{\bf k}\over E_{\bf k}}
\right ),\\
V^{2}_{\bf k}&=&{1\over 2}\left ( 1-{\bar{\varepsilon}_{\bf k}\over E_{\bf k}}
\right ),
\end{eqnarray}
\end{subequations}
with the constraint $U^{2}_{\bf k}+V^{2}_{\bf k}=1$ for any wave vector ${\bf k}$.
The above results in Eqs. (\ref{MF-EGFS}) and (\ref{EBCSCF}) are the standard BCS
expressions for the SC-state\cite{Bardeen57,Schrieffer64}, although the electron
pairing mechanism is driven by the kinetic energy by the exchange of the spin
excitation\cite{Feng15,Feng0306,Feng12,Feng15a}.

Substituting above renormalized electron diagonal and off-diagonal propagators in
Eq. (\ref{MF-EGFS}) and spin propagator in Eq. (\ref{MFSGF}) into Eqs. (\ref{ESE}),
the electron normal and anomalous self-energies now can be derived as,
\begin{widetext}
\begin{subequations}\label{ESE-1}
\begin{eqnarray}
\Sigma_{\rm ph}({\bf{k}},{\omega})&=&\frac{Z_{\rm F}}{N^{2}}
\sum_{{\bf{p}}{\bf{p}'}{\nu}}(-1)^{\nu+1}V^{2}_{\rm ph}
{\Omega}_{{\bf{p}}{\bf{p}'}{\bf{k}}}\left [ U^{2}_{\bf{p}+\bf{k}}\left (
\frac{F_{1\nu}({\bf p},{\bf p}',{\bf k})}{\omega+\omega^{(\nu)}_{{\bf{p}}{\bf{p}}'}
-E_{\bf{p}+\bf{k}}}
-\frac{F_{2\nu}({\bf p},{\bf p}',{\bf k})}{\omega-\omega^{(\nu)}_{{\bf{p}}{\bf{p}}'}
-E_{\bf{p}+\bf{k}}} \right)\right .\nonumber\\
&+&\left .V^{2}_{\bf{p}+\bf{k}} \left (
\frac{F_{1\nu}({\bf p},{\bf p}',{\bf k})}{\omega- \omega^{(\nu)}_{{\bf{p}}{\bf{p}}'}
+E_{\bf{p}+\bf{k}}}
-\frac{F_{2\nu}({\bf p},{\bf p}',{\bf k})}
{\omega+\omega^{(\nu)}_{{\bf{p}}{\bf{p}}'}+E_{\bf{p}+\bf{k}}}  \right )  \right ],
\label{ph-ESE}
\end{eqnarray}
\begin{eqnarray}
\Sigma_{\rm pp}({\bf{k}},{\omega})&=&\frac{Z_{\rm F}}{N^{2}}
\sum_{{\bf{p}}{\bf{p}'}{\nu}}(-1)^{\nu}V^{2}_{\rm pp}
{\Omega}_{{\bf{p}}{\bf{p}'}{\bf{k}}}\frac{\bar{\Delta}_{{\rm{Z}}}({\bf{p}}+{\bf{k}})}
{2E_{{\bf{p}}+\bf{k}}}\left [ \left ( \frac{F_{1\nu}({\bf p},{\bf p}',{\bf k})} {\omega+\omega^{(\nu)}_{{\bf{p}}{\bf{p}}'}-E_{\bf{p}+\bf{k}}}
-\frac{F_{2\nu}({\bf p},{\bf p}',{\bf k})}{\omega-\omega^{(\nu)}_{{\bf{p}}{\bf{p}}'}
-E_{\bf{p}+\bf{k}}} \right)\right .\nonumber\\
&-& \left. \left ( \frac{F_{1\nu}({\bf p},{\bf p}',{\bf k})}
{\omega-\omega^{(\nu)}_{{\bf{p}}{\bf{p}}'}+E_{\bf{p}+\bf{k}}}
-\frac{F_{2\nu}({\bf p},{\bf p}',{\bf{k}})}{\omega+\omega^{(\nu)}_{{\bf{p}}{\bf{p}}'}
+E_{\bf{p}+\bf{k}}}  \right )  \right ], \label{pp-ESE}
\end{eqnarray}
\end{subequations}
respectively, with $\nu=1,2$, and the functions,
\begin{subequations}
\begin{eqnarray}
F_{1\nu}({\bf{p}},{\bf{p}}',{\bf{k}})&=& n_{\rm{F}}(E_{\bf{p}+\bf{k}})
\{1+n_{B}(\omega_{\bf{p}'+\bf{p}})+n_{B}[(-1)^{\nu+1}\omega_{\bf{p}'}]\}
+n_{B}(\omega_{\bf{p}'+\bf{p}})n_{B}[(-1)^{\nu+1}\omega_{\bf{p}'}], \\
F_{2\nu}({\bf{p}},{\bf{p}}',{\bf{k}})&=& [1-n_{\rm{F}}(E_{\bf{p}+\bf{k}})]
\{1+n_{B}(\omega_{\bf{p}'+\bf{p}})+n_{B}[(-1)^{\nu+1}\omega_{\bf{p}'}] \}
+n_{B}(\omega_{\bf{p}'+\bf{p}})n_{B}[(-1)^{\nu+1}\omega_{\bf{p}'}].~~~~
\end{eqnarray}
\end{subequations}

As the case of the charge-carrier normal self-energy in Eq. (\ref{self-energy-2}),
the above electron normal self-energy in Eq. (\ref{ph-ESE}) can be also reexpressed
as,
\begin{eqnarray}\label{EPG-1}
\Sigma_{\rm ph}({\bf k},\omega)&\approx& {[2\bar{\Delta}_{\rm PG}({\bf k})]^{2}
\over \omega -\varepsilon_{\bf k}},
\end{eqnarray}
as quoted in Eq. (\ref{NSPG}) of the main text.


\subsection{Self-consistent equations for determination of electron order parameters}\label{ESCE}

In the framework of the full charge-spin recombination\cite{Feng15a}, the
electron pairing gap and the normal-state pseudogap respectively originate from the
charge-carrier pairing gap and charge-carrier pseudogap. In this case, the electron
pairing gap parameter and normal-state pseudogap parameter can be respectively set as
$\bar{\Delta}=\bar{\Delta}_{\rm h}$ and
$\bar{\Delta}_{\rm PG}=\bar{\Delta}^{\rm (h)}_{\rm pg}$, and then the single-particle
coherent weight, the vertex correction parameters $V_{\rm ph}$ and $V_{\rm pp}$, and
the electron chemical potential satisfy following four self-consistent equations,
\begin{subequations}\label{SCE3}
\begin{eqnarray}
\frac{1}{Z_{\rm{F}}} &=& 1+\frac{Z_{\rm F}}{N^{2}}
\sum_{{\bf{p}}{\bf{p}'}{\nu}}(-1)^{\nu+1}V^{2}_{\rm ph}
\Omega_{{\bf{p}}{\bf{p}'}{\bf k}_{\rm A}}\left (
\frac{F_{1\nu}({\bf p},{\bf p}',{\bf k}_{\rm A})}{(\omega^{(\nu)}_{{\bf{p}}{\bf{p}}'}
-E_{\bf{p}+{\bf k}_{\rm A}})^{2}}
+\frac{F_{2\nu}({\bf{p}},{\bf{p}}',{\bf k}_{\rm A})}
{(\omega^{(\nu)}_{{\bf{p}}{\bf{p}}'}+E_{\bf{p}+{\bf k}_{\rm A}})^{2}} \right ),
~~~~~~~~~\\
1 &=& \frac{4Z^{2}_{\rm F}}{N^{3}}\sum_{{\bf{p}}{\bf{p}'}{\bf{k}}{\nu}}(-1)^{\nu}
V^{2}_{\rm pp}\Omega_{{\bf{p}}{\bf{p}'}{\bf{k}}}\frac{\gamma^{\rm (d)}_{\bf k}
\gamma^{\rm (d)}_{{\bf p}+{\bf k}}}{E_{\bf{p}+\bf{k}}}\left (
\frac{F_{1\nu}({\bf{p}},{\bf{p}}',{\bf{k}})}{\omega^{(\nu)}_{{\bf{p}}{\bf{p}}'}
-E_{\bf{p}+\bf{k}}}-\frac{F_{2\nu}({\bf{p}},{\bf{p}}',{\bf{k}})}
{\omega^{(\nu)}_{{\bf{p}}{\bf{p}}'}+E_{\bf{p}+\bf{k}}} \right ),~~~~~\\
(\bar{\Delta}^{\rm (h)}_{\rm pg})^{2} &=& - \frac{Z_{\rm F}}{4N^{3}}
\sum_{{\bf{k}}{\bf{p}}{\bf{p}'}{\nu}}(-1)^{\nu+1}V^{2}_{\rm ph}\varepsilon_{\bf k}
\Omega_{{\bf{p}}{\bf{p}'}{\bf{k}}}
{\bar{\varepsilon}_{\bf{p}+\bf{k}}\over E_{\bf{p}+{\bf{k}}}}\left (
\frac{F_{1\nu}({\bf p},{\bf p}',{\bf k})}{\omega^{(\nu)}_{{\bf{p}}{\bf{p}}'}
-E_{\bf{p}+\bf{k}}}
+\frac{F_{2\nu}({\bf p},{\bf p}',{\bf k})}{\omega^{(\nu)}_{{\bf{p}}{\bf{p}}'}
+E_{\bf{p}+\bf{k}}} \right),\\
1-\delta &=&\frac{Z_{\rm F}}{2N}\sum_{{\bf{k}}}\left(
1-\frac{\bar{\varepsilon}_{\bf{k}}}{E_{{\bf{k}}}}{\rm{tanh}}
\left[ \frac{1}{2}\beta{E_{\bf{k}}} \right ] \right ),\label{SCE-EFS}
\end{eqnarray}
\end{subequations}
respectively, where ${\bf k}_{\rm A}=[\pi,0]$. Within the same calculation condition
as in the evaluation of the self-consistent equations (\ref{SCE1}) and (\ref{SCE2}),
the above self-consistent equations (\ref{SCE3}) have been also solved simultaneously,
and then the single-particle coherent weight, the vertex correction parameters, and
electron chemical potential are determined {\it without using any adjustable parameters}.
In particular, the evolution of the electron pair gap parameter $\bar{\Delta}$ and
the normal-state pseudogap parameter $\bar{\Delta}_{\rm PG}$ with doping has been
derived, and the results of $\bar{\Delta}$ and $\bar{\Delta}_{\rm PG}$
as a function of doping are plotted in Fig. \ref{EPG-T-doping}a.

\subsection{Electron pair transition temperature}\label{EPTT}

Concomitantly, the evolution of the electron pair transition temperature $T_{\rm c}$
with doping can be also calculated self-consistently from the above self-consistent
equations (\ref{SCE3}) at the condition of $\bar{\Delta}=0$, and the
result\cite{Feng15a} of $T_{\rm c}$ as a function of doping is plotted in
Fig. \ref{EPG-T-doping}b (black-line), where in a given doping, the magnitude of
$T_{\rm c}$ is exactly identical to the corresponding magnitude\cite{Feng15,Feng12}
of the charge-carrier pair transition temperature $T^{\rm (pair)}_{\rm c}$. This
follows a basic fact that in the framework of the kinetic-energy-driven
superconductivity\cite{Feng15,Feng0306,Feng12,Feng15a}, the effective attractive
interaction between charge carriers originates in their coupling to the spin
excitation. On the other hand, the electron pairing interaction in the framework of
the full charge-spin recombination\cite{Feng15a} is mediated by the exchange of the
same spin excitation, which thus leads to that the magnitude of $T_{\rm c}$ is the
same as the magnitude of the corresponding $T^{\rm (pair)}_{\rm c}$, and then the
dome-like shape of the
doping dependence of $T_{\rm c}$ with its maximum occurring at around the optimal
doping is a natural consequence of the dome-like shape of the doping dependence of
$T^{\rm (pair)}_{\rm c}$ with its maximum occurring at around the same optimal doping.

\subsection{Normal-state pseudogap crossover temperature}\label{NSPCT}

As in the case of the temperature dependence of the electron pair gap $\bar{\Delta}$,
the normal-state pseudogap $\bar{\Delta}_{\rm PG}$ is also strongly temperature
dependent. In particular, in a given doping, the normal-state pseudogap vanishes when
temperature reaches the normal-state pseudogap crossover temperature $T^{*}$. In this
case, we have also made a series of calculations for $T^{*}$ at different doping
concentrations, and the results of $T^{*}$ as a function of doping is plotted in
Fig. \ref{EPG-T-doping}b (red-line).

\end{widetext}

\end{appendix}

\end{document}